 \pdfoutput=1
\newcommand*{\ATLASLATEXPATH}{latex/}
\documentclass[cernpreprint,txfonts,UKenglish,texlive=2016,titleextraheight=5em]{\ATLASLATEXPATH atlasdoc}  %

\usepackage[subfigure]{\ATLASLATEXPATH atlaspackage}
\usepackage{\ATLASLATEXPATH atlasbiblatex}

\usepackage{\ATLASLATEXPATH atlascontribute}

\usepackage{\ATLASLATEXPATH atlasphysics}

\graphicspath{{logos/}{figures/}}

\usepackage{lvqqpaper-defs}

\usepackage{dcolumn}
\newcolumntype{?}{D{,}{\,\pm\,}{-1}}
\newcolumntype{!}{D{,}{\,\pm\,}{4.4}}

\AtlasTitle{Measurement of $\WWWZ$ production with the hadronically decaying 
boson reconstructed as one or two jets in $pp$ collisions at $\rts=8~\tev$ 
with ATLAS, and 
constraints on anomalous gauge couplings}

\author{The ATLAS Collaboration}

\AtlasRefCode{STDM-2015-23}

\PreprintIdNumber{CERN-EP-2017-060}

 \arXivId{1706.01702}

 \AtlasJournalRef{\EPJC 77 (2017) 563}
 \AtlasDOI{doi:10.1140/epjc/s10052-017-5084-2}

\AtlasAbstract{%
This paper presents a study of the production of $WW$ or $WZ$ boson 
pairs, with one $W$ boson decaying to $e\nu$ or $\mu\nu$ and one
$W$ or $Z$ boson decaying hadronically.  The analysis uses $\ourLumiNoError$ 
of $\rts=8~\tev$ $pp$ collision data, collected by the ATLAS detector at
the Large Hadron Collider.  Cross-sections for  $WW/WZ$ production
are measured in high-$\pT$ fiducial regions defined close 
to the experimental
event selection. The cross-section is measured 
for the case where the hadronically decaying boson is 
reconstructed as two resolved jets, and the case where it is reconstructed as a single jet.
The transverse momentum distribution of the hadronically decaying
boson is used to search for new physics.
Observations are consistent with the Standard Model predictions, and
$95\%$ confidence intervals are calculated for parameters describing
anomalous triple gauge-boson couplings.
}

\hypersetup{pdftitle={ATLAS document},pdfauthor={The ATLAS Collaboration}}

\begin{document}

\maketitle

\tableofcontents

%-------------------------------------------------------------------------------
\section{Introduction}
\label{sec:intro}

Measurements of the production of two massive vector gauge bosons
(hereafter, ``diboson'' production) represent an important 
test of the Standard Model (SM) of particle physics.  
Diboson measurements are powerful probes of the electroweak
theory of the SM, in particular the structure of the
triple gauge-boson couplings (TGCs)~\cite{Hagiwara:1986vm,Butter:2016cvz}.  
In addition, precise
diboson measurements are a valuable test of higher-order
calculations in quantum chromodynamics (QCD).

Measurements of $WW$  and $WZ$ production 
in the leptonic channels $\ell\nu\ell\nu$ 
and $\ell\nu\ell\ell$ ($\ell=e,\mu$)
have been performed by the
ATLAS and CMS collaborations 
in $pp$
collisions at $\rts=8~\tev$  
and $\rts=13~\tev$~\cite{STDM-2013-07,Aaboud:2017qkn,CMS-SMP-14-016,STDM-2014-02,STDM-2015-19,Khachatryan:2016poo,Khachatryan:2016tgp},
and by the Tevatron experiments in $p\bar{p}$ collisions~\cite{Abazov:2009ys,Aaltonen:2009aa,Aaltonen:2012vu,Abazov:2012cj}.
Measurements in the semileptonic channel $\WVlnuqq$ ($V=W,Z$) have been performed
by ATLAS~\cite{STDM-2012-22} and CMS~\cite{CMS-SMP-12-015} at $\rts=7~\tev$,
and by the Tevatron experiments 
in $p\bar{p}$ collisions~\cite{Abazov:2011cb,Aaltonen:2016pos}.
The semileptonic channel offers features complementary 
to the leptonic channels.  On the one hand, the presence of jets
and the large background from $\Wjet$ and $\ttbar$ production 
limit the experimental precision.     
On the other hand, the semileptonic channel has an approximately six times higher branching
fraction than the fully leptonic channels.  Also, for $WW$, the original diboson
kinematics can be better reconstructed in an $\lnuqq$ final state
than in an $\lnulnu$ final state, since the latter has two invisible
particles, rather than only one in $\lnuqq$.  Both of these advantages
are particularly beneficial for searching for beyond-the-Standard-Model (BSM) enhancements of
diboson production due to heavy new particles, which could modify
the diboson spectrum at high transverse momentum ($\pT$) of
the bosons~\cite{Falkowski:2016cxu}.

It is possible to reconstruct the $\Vqq$ decay
as two small-radius jets (``small-$R$'' jets, denoted by $\jet$)
or as a single large-radius jet (``large-$R$'' jet, denoted by $\fatjet$).
Reconstructing the $\Vqq$ decay as a large-$R$ jet
enables an increased reconstruction
efficiency at high $\pT(V)$, thus improving the sensitivity to
BSM signals.
 In addition, by applying grooming~\cite{Altheimer:2013yza} techniques 
such as trimming~\cite{Krohn:2009th} 
to the large-$R$ 
jets, it is possible to better distinguish events
containing $\Vqq$ decays from background events~\cite{PERF-2015-03}.

In this paper, measurements of $\WVlnuqq$ fiducial
cross-sections are presented in phase spaces containing a $\Vqq$ 
candidate with high $\pT$. Two fiducial
cross-sections are measured, in phase spaces chosen to closely 
match the two 
experimental selections used in this paper.  The first event selection,
denoted \resolved,
reconstructs the $\Vqq$ decay as two small-$R$ jets, while the second 
one, denoted \boosted, reconstructs
the $\Vqq$ as a single large-$R$ jet.
Previous cross-section measurements of
 $\WVlnuqq$ have not exploited large-$R$ jets.

A search for anomalous triple gauge-boson couplings (aTGCs) is also presented in 
this paper, using both the \resolved\ and \boosted\ channels.  
Previous searches for charged aTGC contributions
to $\WVlnuqq$ production have been conducted by the ATLAS
Collaboration~\cite{STDM-2012-22} using $7~\tev$ $pp$ collisions, by
the CMS Collaboration~\cite{CMS-SMP-12-015,Sirunyan:2017bey} using $7$ and $8~\tev$ $pp$ collisions,
and by the
D0~\cite{Abazov:2012ze} and CDF~\cite{Aaltonen:2007sd} collaborations using 
$\ppbar$ collisions.
Most published aTGC searches 
in the $\WVlnuqq$ channel have reconstructed the $\Vqq$ as two small-$R$ jets,
with the exception of Ref.~\cite{Sirunyan:2017bey}, which
reconstructed the $\Vqq$ as a single large-$R$ jet.

%-------------------------------------------------------------------------------

%-------------------------------------------------------------------------------
\section{Analysis overview}
\label{sec:overview}
As mentioned above, measurements of $\WVlnuqq$ production are
performed using either two small-$R$ jets or a single
large-$R$ jet to reconstruct the hadronically decaying $V$
boson.
For both channels, the leptonically decaying $W$ boson is reconstructed 
by requiring the presence of
a lepton (electron or muon) and missing transverse momentum.
 
After applying stringent event selection requirements, the signal-to-background ratio remains quite low at 5--10$\%$, because of the large W + jets background. 
In order to distinguish
the SM $\WV$ signal from the background, the 
dijet mass distribution (in the
\resolved\ channel) or the mass distribution of the large-$R$ jet 
(in the \boosted\ channel) is used as a discriminating variable.
The signal events peak near the $W/Z$ mass in these distributions, while
the shape of the dominant $\Wjet$ background is smoothly falling.  
In both
channels, the signal is extracted from a fit to the discriminating variable.
Wide fitting ranges are used, in order to allow the backgrounds to 
be constrained by the data.

A fiducial cross-section is measured separately in the
 \resolved\ and the \boosted\ channel; the 
fiducial phase spaces for the measurements are defined to be close to the
experimental event selections.  
The fiducial cross-section in each channel is extracted from 
the previously mentioned fits. 
The events in the two channels partially
overlap, because there are some events for which the $\Vqq$ decay 
can be reconstructed both as two small-$R$ jets and as one large-$R$ jet.
In order to simplify 
the interpretation of the results and allow
easier comparison with theoretical predictions, the overlap events are
not removed, and both measurements are presented separately.
No combination of the \resolved\ and \boosted\
cross-section measurements is performed.
The electron and muon channels are combined when performing the measurements,
since little improvement in sensitivity 
is expected from separating by lepton flavour. 
Event kinematics
and the signal-to-background ratio are similar in the electron and muon channels,
and the dominant sources of uncertainty are unrelated to lepton flavour.

A search for aTGC contributions is also performed in the
\resolved\ and \boosted\ channels.  The event selection is
the same as for the cross-section measurements, except that a tighter
requirement is made on the dijet mass or on the mass of the large-$R$ jet.
The search is performed by fitting the $\pT$
distribution of the dijet system  (\resolved\
channel) or of the large-$R$ jet (\boosted\ channel). These
distributions are sensitive to aTGCs, which are expected
to lead to deviations from the SM prediction at high $\pT$.

%-------------------------------------------------------------------------------

%-------------------------------------------------------------------------------
\section{ATLAS detector}
\label{sec:detector}
The ATLAS detector~\cite{PERF-2007-01}, which
surrounds one of the interaction points of the Large Hadron Collider (LHC)~\cite{Evans:2008zzb}, is built of several subdetectors.
The first subdetector layer consists of the inner detector (ID), 
which provides charged-particle tracking for $|\eta|<2.5$.\footnote{ATLAS uses a right-handed coordinate system with its origin at the nominal interaction point (IP) in the centre of the detector and the $z$-axis along the beam pipe. The $x$-axis points from the IP to the centre of the LHC ring, and the $y$-axis points upward. Cylindrical coordinates $(r,\phi)$ are used in the transverse plane, $\phi$ being the azimuthal angle around the $z$-axis. The pseudorapidity is defined in terms of the polar angle $\theta$ as $\eta=-\ln\tan(\theta/2)$.}
  The ID is 
further subdivided into (ordered from innermost to outermost) 
a pixel detector, a silicon-microstrip tracker,
and a transition radiation tracker.  Surrounding the ID there is a
superconducting solenoid that provides a 2~T magnetic field.  Outside
of the solenoid, there is an electromagnetic (EM) calorimeter based on 
liquid-argon technology, which provides coverage up to $|\eta|=3.2$. 
Additionally, a scintillator-tile calorimeter provides hadronic energy
measurements in the range $|\eta|<1.7$, and liquid-argon-based endcap and
forward calorimeters extend the EM and hadronic measurements up to $|\eta|= 4.9$.
A muon spectrometer, consisting of tracking and triggering detectors
and three toroidal magnets,
surrounds the calorimeters; it provides muon
tracking and identification up to $|\eta|=2.7$ and triggering 
capability up to $|\eta|=2.4$.

A three-level trigger
system is used to select the most interesting events for data
storage~\cite{PERF-2011-02}. An initial hardware-based 
trigger stage is followed by
two software-based triggers, which reduce the final event rate
to about 400 Hz.

%-------------------------------------------------------------------------------

%-------------------------------------------------------------------------------
\section{Data and Monte Carlo samples}
\label{sec:samples}
This analysis is based on an integrated luminosity of $\ourLumi$ of $8~\tev$ $pp$ collisions recorded
by the ATLAS detector in 2012.  Events are required to pass 
one of several single-lepton triggers.  The triggers require 
either an isolated electron or muon with $\pT>24~\gev$, or 
an electron (muon) having $\pT>60~(36)~\gev$ without an isolation
requirement.

The nominal signal Monte Carlo (MC) samples consist of  $\qqtoWV$ events generated
at next-to-leading order (NLO) in QCD  using 
\MCatNLO~v4.07~\cite{Frixione:2002ik} interfaced 
with \HERWIG~v6.520~\cite{Corcella:2000bw} 
and \JIMMY~v4.31~\cite{Butterworth:1996zw} for the simulation of
parton showering, hadronization, and the underlying event.  
The CT10 parton distribution function (PDF) set~\cite{Lai:2010vv} 
and parameter values from the AUET2 tune~\cite{ATL-PHYS-PUB-2011-008} 
are used for these samples.  
The $W$ and $Z$ bosons
are generated on-shell by \MCatNLO and decayed subsequently by
\HERWIG. The same MC configuration is also used to model aTGC
contributions to $WV$ production, using an event reweighting
feature built into \MCatNLO.

In order to study systematic uncertainties, 
alternative $\qqtoWV$ samples are generated at NLO in QCD
with \POWHEGBOX~\cite{Nason:2004rx,Frixione:2007vw,Alioli:2010xd}
using the CT10 PDF set.
The parton showering and hadronization is modelled with \PYTHIA~8.175~\cite{Sjostrand:2007gs}
using the AU2 tune~\cite{ATL-PHYS-PUB-2012-003}.
Off-shell $W$ and $Z/\gamma^{*}$ decays are included; the  $Z/\gamma^{*}$ decays 
have a requirement of $m_{\qq}>20~\gev$ and $m_{\ell\ell}>20~\gev$.

Another set of alternative $\qqtoWV$ samples are generated 
with \SHERPA~v1.4.1~\cite{Gleisberg:2008ta,Hoeche:2009rj,Gleisberg:2008fv,Schumann:2007mg}.  
These samples are generated at leading order (LO) in QCD, but
include up to three additional partons in the matrix element.
Off-shell $W$ and $Z/\gamma^{*}$ decays are included; the  $Z/\gamma^{*}$ decays 
have a requirement of  $m_{\qq}>4~\gev$ and $m_{\ell\ell}>4~\gev$.

Contributions from $gg\to H\to WW^{*}$ are only at the $1\%$ level after applying the full
event selection and are thus neglected. Signal MC
samples for non-resonant $\ggtoWW$ production are not used
in the analysis, but the contribution from this process
is estimated as described
in Section~\ref{sec:xsecResult}, and included in the final cross-section 
predictions.

The $\Wjet$ and $\Zjet$ backgrounds (collectively referred to as $\Vjet$) are 
modelled at LO in QCD with \SHERPA~v1.4.1,
with up to four additional final-state partons. 
The CT10 PDF set is
used for these samples, and they are normalized using
inclusive cross-sections that are next-to-next-to-leading order (NNLO) in
QCD, obtained using FEWZ~\cite{Anastasiou:2003ds}.  
For studies of systematic uncertainties,
alternative $\Wjet$ samples are generated with \ALPGEN~\cite{Mangano:2002ea} interfaced with \PYTHIA~6.426~\cite{Sjostrand:2006za},  
modelling the process at LO in QCD with up to five final-state partons. 
These additional samples use the Perugia 2011C tune~\cite{Skands:2010ak} and the CTEQ6L1 PDF set~\cite{Pumplin:2002vw}.

The MC samples for the \ttbar\ and 
single-top-quark ($t$-channel, $s$-channel, and $Wt$) processes
(collectively referred to as top-quark processes)
are generated with \POWHEGBOX~\cite{
Frixione:2007nw,Alioli:2009je,Re:2010bp} 
interfaced with \PYTHIA~6.426~\cite{Sjostrand:2006za}
(or \PYTHIA~6.427 for the $t$-channel single-top-quark process).  All of
these samples use the CT10 PDF set for the matrix element, the 
CTEQ6L1 PDF set for the parton shower, and the  Perugia 2011C tune.

The $ZZ$ background process is modelled with 
\POWHEG\ interfaced with \PYTHIA~8.
The sample is normalized using the NLO prediction from \MCFM~\cite{Campbell:2010ff,Campbell:1999ah}.

The MC samples are passed through a GEANT4-based~\cite{GEANT4} simulation 
of the ATLAS detector~\cite{SOFT-2010-01}.  For some of the
MC samples, a fast simulation is used that makes use of a
parameterization of the showers in the calorimeter.
The hard-scattering processes in the MC samples are overlaid with
simulated minimum-bias events in order to model additional collisions
in the same or neighbouring bunch crossings (``pile-up'').
The MC samples are reweighted so that their pile-up profile matches
that observed in the data.

%-------------------------------------------------------------------------------

%-------------------------------------------------------------------------------
\section{Event reconstruction}
\label{sec:objselection}
This analysis considers events with exactly one lepton (electron or muon), missing transverse momentum, and either two small-$R$ jets or
one large-$R$ jet.

In each event, primary vertices are reconstructed, which must be formed from at 
least three tracks with $\pT>400~\mev$.  In case an event has multiple primary
vertices (due to pile-up), the primary vertex with the highest $\sum \pT^2$ of the
associated tracks is defined as the hard-scatter vertex.

Electron candidates are formed from energy clusters in the 
EM calorimeter matched to ID tracks.  They are required
to have $\pT>30~\gev$ and $|\eta|<2.47$. Candidates
in the transition region between the barrel and endcaps of the
EM calorimeter, $1.37<|\eta|<1.52$, are excluded.  
In order to ensure that the electron candidates 
are consistent with having been  produced at the
hard-scatter vertex, 
the transverse
impact parameter $d_0$ and longitudinal impact parameter
$z_0$ are required to satisfy $\dzsig<5$ and $|z_0 \sin{\theta}|<0.5$~mm,
respectively, where $\sigmadz$ is the uncertainty in 
the measured $d_0$.  Both $d_0$ and $z_0$ are measured with respect
to the hard-scatter vertex.  Electron candidates must also satisfy
the ``tight'' cut-based identification criteria 
from Ref.~\cite{Aaboud:2016vfy},  
based on track parameters and on the shower shapes in the calorimeter.
Candidates must also pass isolation requirements based
on calorimeter and track measurements. The calorimeter
isolation requires
$\caloiso<0.14$, where 
$\caloiso$ is
defined as the scalar transverse energy sum of the 
 calorimeter energy deposits 
within a $\DR\equiv\sqrt{(\Delta\eta)^2 + (\Delta\phi)^2}=0.3$ cone
centred on the electron candidate (excluding transverse energy from the candidate itself),
divided by the $\pT$ of the electron candidate.
Similarly, the track isolation requires
$\trackiso<0.07$,  where $\trackiso$
is the scalar sum of the $\pT$ of the tracks
within a $\DR=0.3$ cone
centred on the electron candidate (excluding
the $\pT$ of the candidate's track itself), divided by the electron candidate's $\pT$.

Muon candidates are formed from the combination of a
track in the muon spectrometer and one in the ID.  They are required
to have $\pT>30~\gev$ and $|\eta|<2.4$.  Their impact parameters
must satisfy  $\dzsig<3$ and $|z_0 \sin{\theta}|<0.5$~mm.  
The candidates must also satisfy the isolation criteria
$\caloiso<0.07$ and $\trackiso<0.07$, where $\caloiso$ 
and $\trackiso$ are defined analogously to the electron case.

Small-$R$ jets are reconstructed from topological 
energy clusters~\cite{ATL-LARG-PUB-2008-002} in the calorimeter using the
anti-$k_t$ algorithm~\cite{Cacciari:2008gp} with radius parameter $R=0.4$.
The jet energies are calibrated as described in Ref.~\cite{PERF-2012-01} and are corrected
for pile-up.
They are required to have $\pT>25~\gev$ and $|\eta|<2.5$
for the \resolved\ channel.  Small-$R$ jets with $|\eta|<4.5$ 
are used in the \boosted\ channel as part of a jet veto 
(see Section~\ref{sec:eventSelection}).  
 In order to remove jets originating from pile-up,
small-$R$ jets having $\pT<50~\gev$
and $|\eta|<2.4$ are required
to have an absolute value of the 
``jet vertex fraction'' variable (JVF)~\cite{Aad:2015ina} 
 greater than 0.5.

In the \boosted\ channel, large-$R$ jets are 
reconstructed using the anti-$k_t$ algorithm with radius
parameter $R=1.0$, and are trimmed~\cite{Krohn:2009th} using a subjet radius
of $0.2$ and a momentum-fraction parameter $f_\mathrm{cut}=0.05$; the trimming
procedure discards soft subjets from the large-$R$ jets and reduces their sensitivity
to pile-up~\cite{PERF-2015-03}.
They are required to have $\pT>200~\gev$ and $|\eta|<2.0$. 
The energies of the small-$R$ and large-$R$ jets and the masses of
the large-$R$ jets are calibrated using $\pT$- and $\eta$-dependent 
scale factors~\cite{PERF-2012-01,PERF-2012-02}.

If an 
electron and a muon candidate share the same ID track, the electron 
candidate is rejected.  If a small-$R$ jet is within $\DR=0.2$
of a selected electron candidate, the jet is rejected; if the jet is within
$0.2<\DR<0.4$ of a selected electron, the electron candidate is
rejected.  Muon candidates are rejected if they are within
$\DR=0.4$ of a small-$R$ jet.  Finally, large-$R$ jets are rejected if
they are within $\DR=1.0$ of a selected lepton candidate.
In the object selection stage, small-$R$ jets and large-$R$ jets
are allowed to overlap; however, in the event selection stage
a $\DR$ requirement is applied between the small-$R$ and large-$R$ jets,
as explained in Section~\ref{sec:eventSelection}.

The missing transverse momentum $\vecmet$ is computed
as the negative vector sum of the transverse momentum of all the detected objects in 
the event, including reconstructed jets, photons, electrons, and muons.
An additional ``soft term'' is  included that accounts
for the  $\pT$ of clusters in the calorimeter which are not 
associated with any 
specific reconstructed object~\cite{PERF-2014-04}. The magnitude
of $\vecmet$ is denoted $\met$.

%-------------------------------------------------------------------------------

\section{Event selection}
\label{sec:eventSelection}
Two independent sets of event selection criteria 
are developed that target different 
event topologies:  the \resolved\ selection, 
described in Section~\ref{eventSelectResolved},
and the \boosted\ selection, described in Section~\ref{eventSelectBoosted}.
The \boosted\ channel and \resolved\ channel
differ significantly from one another in their
kinematics, expected signal yields, and signal-to-background ratios.  Therefore,
the event selection criteria are optimized separately for the two channels.

For both the \resolved\ and \boosted\ selections, 
all events are required to contain at least one primary vertex.
Events must have exactly one good electron or muon candidate. 
Events are vetoed if they contain any additional lepton candidates
that have $\pT>15~\gev$ and satisfy a looser set of selection criteria.

\subsection{\resolved\ channel}
\label{eventSelectResolved}
Events must have $\MET>40~\gev$ and a 
transverse mass\footnote{The transverse mass is defined as $\MT \equiv \sqrt{(\MET+\pT(\ell))^2 - |\vecmet+\vec{\pT}(\ell)|^2}$, 
where $\pT(\ell)$ is the transverse momentum of the lepton candidate.} $\MT>40~\gev$. 
Events must contain exactly two small-$R$ jets.
The requirement of exactly two jets substantially
reduces the background from top-quark decays. The pseudorapidity separation
of the selected jets is required to satisfy
$\Delta\eta(\jet,\jet)<1.5$, in order to improve the signal-to-background
ratio.

In order to reduce the multijet background
not removed by the $\met>40~\gev$ requirement,
an azimuthal-angle difference between the $\MET$ direction and the
direction of the leading-\pt jet of \linebreak $|\Delta\phi(\jet_1,\MET)| > 0.8$
is required.
 Also, both the $\Vqq$ and $\Wlv$ candidates must pass requirements on 
their transverse
momenta: $\pTjj>100~\gev$ and $\pTWlep>100~\gev$, where
$\pTWlep\equiv |\vecmet+\vec{\pT}(\ell)|$.  
These $\pT$ requirements enhance the separation between the signal and background
distributions in the dijet mass.

As described in Section~\ref{sec:xsecMethod}, the signal is extracted 
using a maximum-likelihood (ML) fit
to the dijet mass ($\mjjless$) distribution. In the dijet mass calculation,
the mass of each individual jet is set to zero, which makes the
variable easier to model in the MC simulation.
Since the signal is extracted from a fit to $\mjjless$, only a loose
requirement is made on this variable: $40~\gev<\mjjless<200~\gev$.

\subsection{\boosted\ channel}
\label{eventSelectBoosted}
Events must contain exactly one large-$R$ jet  with $\pT>200~\gev$ 
and $|\eta|<2.0$.
The backgrounds from top-quark decays are suppressed by rejecting events containing
any  small-$R$ jets with $\pT>25~\gev$ and $|\eta|<4.5$ that are separated
from the large-$R$ jet by $\DR(\jet,\fatjet)>1.0$.
In order to suppress the multijet background, a requirement 
of $\met>50~\gev$ is applied.
The trimmed mass of the  large-$R$ jet, $\mJ$, must  
be $50~\gev<\mJ<170~\gev$, and the signal
is measured from the ML fit to $\mJ$.

Since the \resolved\ and \boosted\ event selections are done independently, 
some events pass both selections.  
About $10\%$ of the signal MC events that pass the \resolved\ selection also 
pass the \boosted\ selection, while about $50\%$ of the signal MC
events that pass the \boosted\ selection
also pass the \resolved\ selection.

%-------------------------------------------------------------------------------

%-------------------------------------------------------------------------------
\section{Background estimation}
\label{sec:backgrounds}
The methods for estimating the expected background
yields and kinematic distributions are described in this
section.
The estimates from this section are used as inputs to the ML fit in which the signal is measured
while the backgrounds are allowed to vary within their systematic uncertainties.
In that ML fit, the $\Vjet$ normalization is allowed to vary without constraint,
so the estimates given in this section are pre-fit
estimates.

Most of the backgrounds ($\Wjet$, $\Zjet$, $\ttbar$, single top-quark, and $ZZ$) are estimated using
MC simulation, with data-driven corrections applied in some cases, as described later in this section.  
By far the largest background in the analysis is from $\Wjet$, followed by
top-quark production.  Despite the latter background's subdominant contribution, it plays an important
role because it contains contributions from real $\Wtoqq$ decays, which make it more difficult to 
distinguish from the signal.  About $80\%$ of the top-quark background is due to $\ttbar$ production,
and the remainder comes from single-top-quark production.

Multijet processes form another source of background.  Multijet events can
pass the event selection if they contain non-prompt leptons
(produced from semileptonic decays of $c$- and $b$-hadrons) or ``fake'' leptons 
(resulting from  misidentified jets).
The multijet backgrounds are estimated using data-driven techniques, as described in Sections~\ref{sec:backgrounds_resolved} and
\ref{sec:backgrounds_boosted}.

\subsection{\resolved\ channel}
\label{sec:backgrounds_resolved}
\begin{figure}[htbp]
\begin{center}

\subfigure[]{
 \includegraphics[width=0.47\textwidth,keepaspectratio]{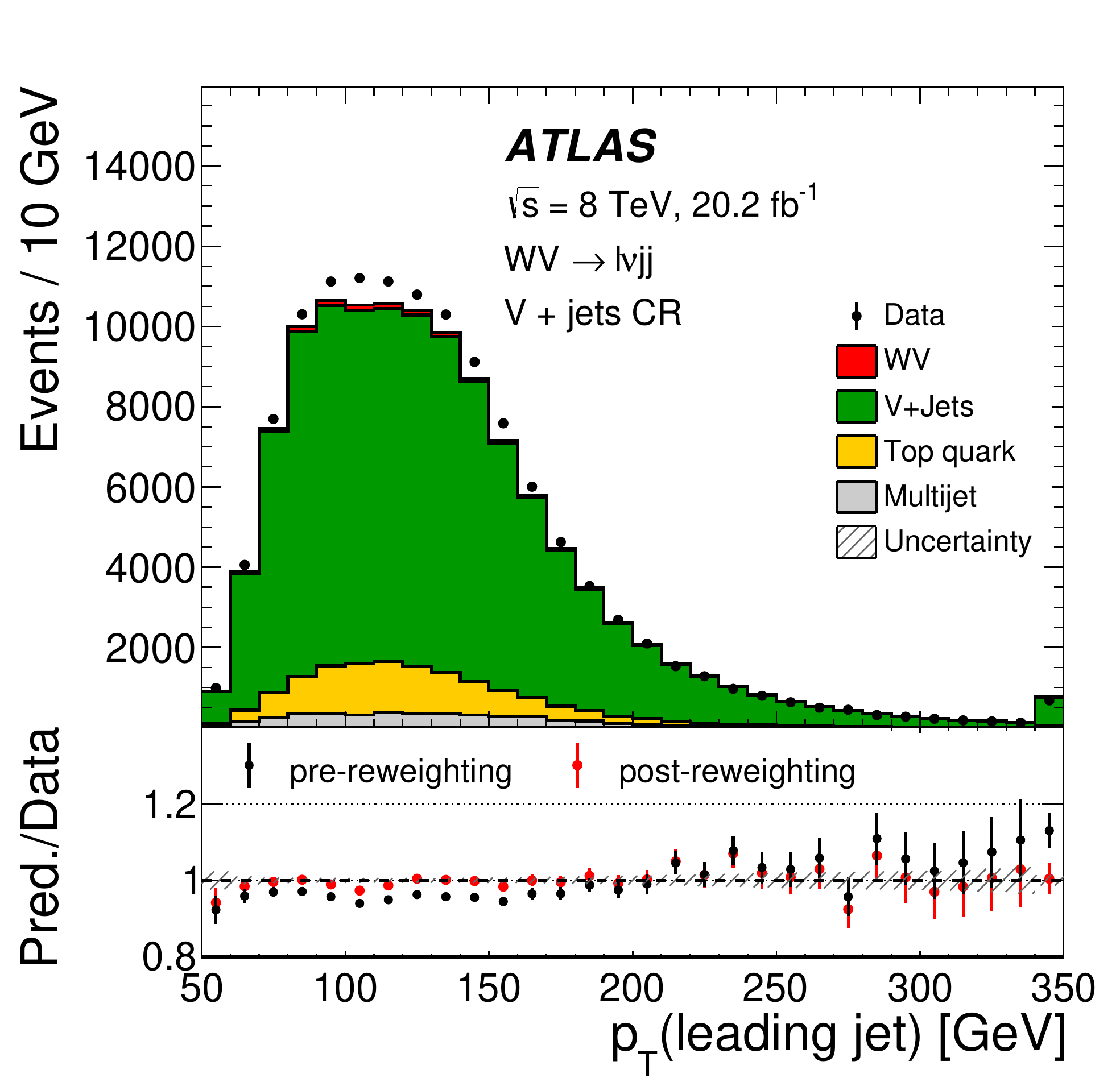}
}
\subfigure[]{
 \includegraphics[width=0.47\textwidth,keepaspectratio]{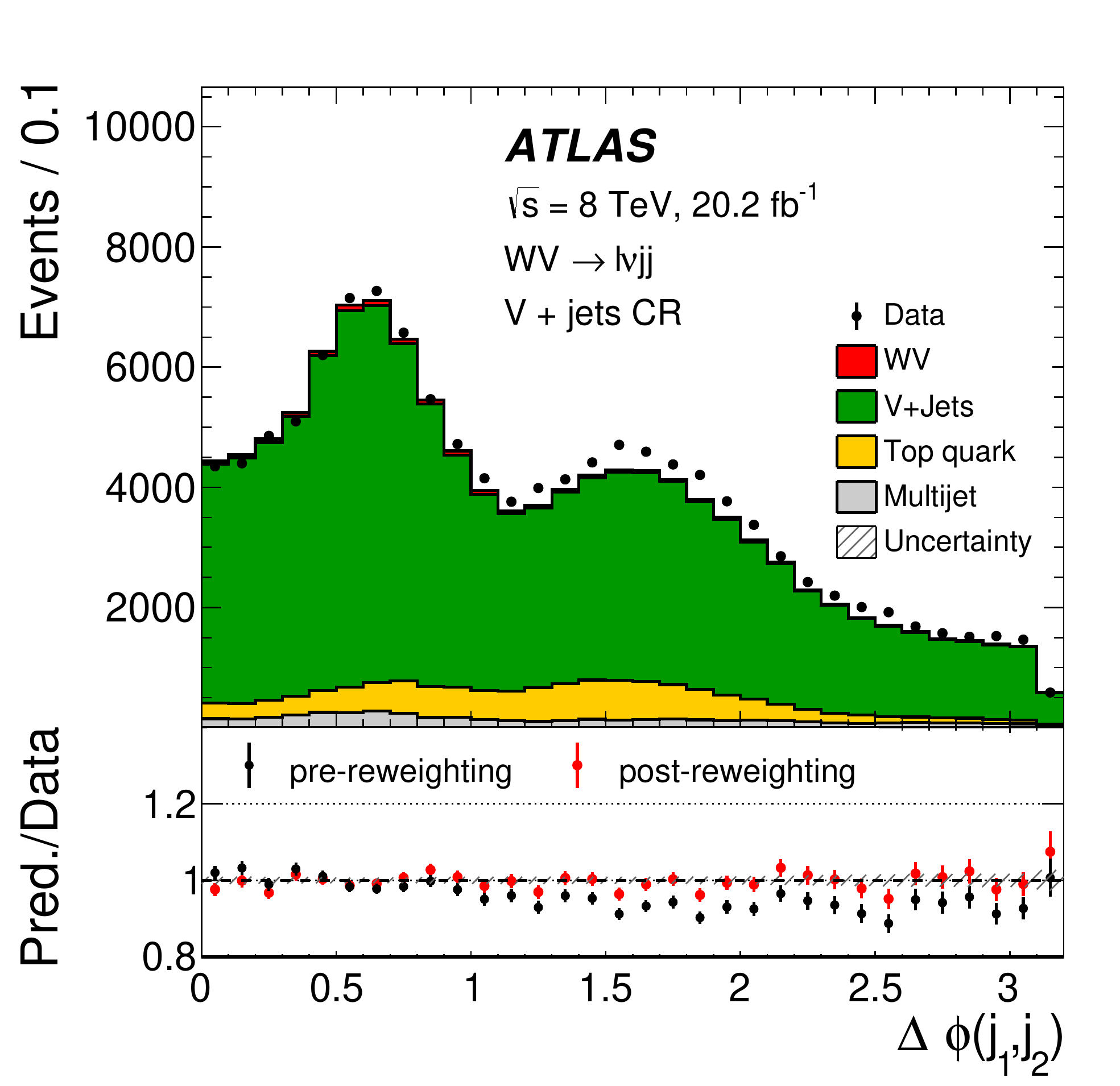}
}
\caption{
Comparisons between the data and the prediction in the $\Vjet$ control region of the \resolved\ channel.
The top panel shows the data and prediction
before applying the $\pT(\jet_1)$ and $\Delta\phi(\jet_1,\jet_2)$ kinematic reweighting 
to the $\Vjet$ predictions.
The distributions shown are
(a) $\pT$ of the leading jet and
(b) $\Delta \phi$ between the leading jet and sub-leading jet. 
Overflow is included in the last bin of the  $\pT(\jet_1)$ plot. 
The bottom panel shows the ratio of the SM prediction to the data before and after 
applying the kinematic reweighting to the $\Vjet$ prediction.
The hatched bands indicate the statistical uncertainty in the predictions. 
}
\label{fig:prefit_dataMC_resolved_reweighting}
\end{center}
\end{figure}

The $\Vjet$ background prediction is MC-based, but data-driven corrections are applied
to the MC prediction in order to improve the description of the jet kinematics.
A  $\Vjet$ control region (CR) is defined identically to the signal region, except
that the region $65~\gev<\mjjless<95~\gev$ is vetoed, in order to remove most of the signal events.
One-dimensional reweighting functions of the variables $\pT(\jet_1)$ and $\Delta\phi(\jet\jet)$
are derived from this $\Vjet$ CR. 
These reweighting functions have approximately $10\%$ effects on the shapes of the $\pT(\jet_1)$ and $\Delta\phi(\jet\jet)$ 
 distributions. Data--MC comparisons in the $\Vjet$ CR are shown in Figure~\ref{fig:prefit_dataMC_resolved_reweighting}, 
before and after application of the reweighting functions.
All further results in this paper are shown with these two reweighting functions applied to the $\Vjet$
MC samples.  The same reweighting functions are used for both the $\Wjet$ and $\Zjet$ processes.
It was checked that the reweighting functions obtained from the low-$\mjjless$ and high-$\mjjless$
portions of the $\Vjet$ control region are compatible.

The top-quark background is modelled with MC simulation, and is cross-checked in a
validation region containing three small-$R$ jets, one of which is 
 $b$-tagged using the MV1 algorithm~\cite{ATLAS-CONF-2014-046,Aad:2015ydr}.
Good agreement is observed between the data and the MC simulation, so no corrections are applied to the prediction.
The background from $ZZ$ events is also modelled with MC simulation.

The data-driven multijet background estimate makes use of a multijet CR.  The multijet CR
is formed by selecting events in data that pass the same selection requirements as for the signal
region, except that the lepton quality criteria are modified in order to produce a CR enriched
in non-prompt and fake leptons. Lepton candidates satisfying these modified criteria are called
``anti-identified'' lepton candidates.  Anti-identified muon candidates must  have a non-negligible impact 
parameter, $\dzsig>4$, and satisfy looser isolation criteria than the signal muon candidates.
Anti-identified electrons must fail the ``tight'' but satisfy the ``medium'' cut-based 
identification criteria from Ref.~\cite{Aaboud:2016vfy}, and are also 
required to contain a hit in the
innermost layer of the pixel detector.  In addition, the isolation criteria are modified
for anti-identified electron candidates, in order to enrich the sample in non-prompt and fake electrons.

The shapes of the kinematic distributions (such as $\mjj$, $\met$, and $\pTjj$) of the 
multijet background are estimated from events in the multijet CR, after 
subtracting the MC predictions of the non-multijet contributions to the CR.  These non-multijet
contributions are about 20\% (50\%) of the total in the electron (muon) channel.
The overall multijet background
event yield is estimated from a fit to the $\met$ distribution of events that
pass the full signal region selection, except that the requirements on $\met$ and
$\Delta\phi(\jet_1,\MET)$ (and also $\Delta\eta(\jet,\jet)$ and $\MT$ for the muon channel)
are  removed in order to enhance the number of multijet events. This selection is referred to
as the \textit{extended} signal region.  In this $\met$ fit, the multijet $\met$ shape is estimated from an
extended multijet CR, defined analogously to the extended signal region, but requiring the lepton 
to pass the anti-identified-lepton
selection.  The $\met$ shapes of the other backgrounds are estimated using MC samples.
The multijet event yield obtained from this fit is then extrapolated to the signal region, using the ratio
of events in the multijet CR and the extended multijet CR, corrected for non-multijet contributions.
The multijet background estimates are performed separately for the electron and muon
channels.  Only about $5\%$ of the total multijet background is in the
muon channel.

\begin{figure}[htbp]
\begin{center}
\subfigure[]{
  \label{fig:templates_normalized_resolved}
 \includegraphics[width=0.47\textwidth,keepaspectratio]{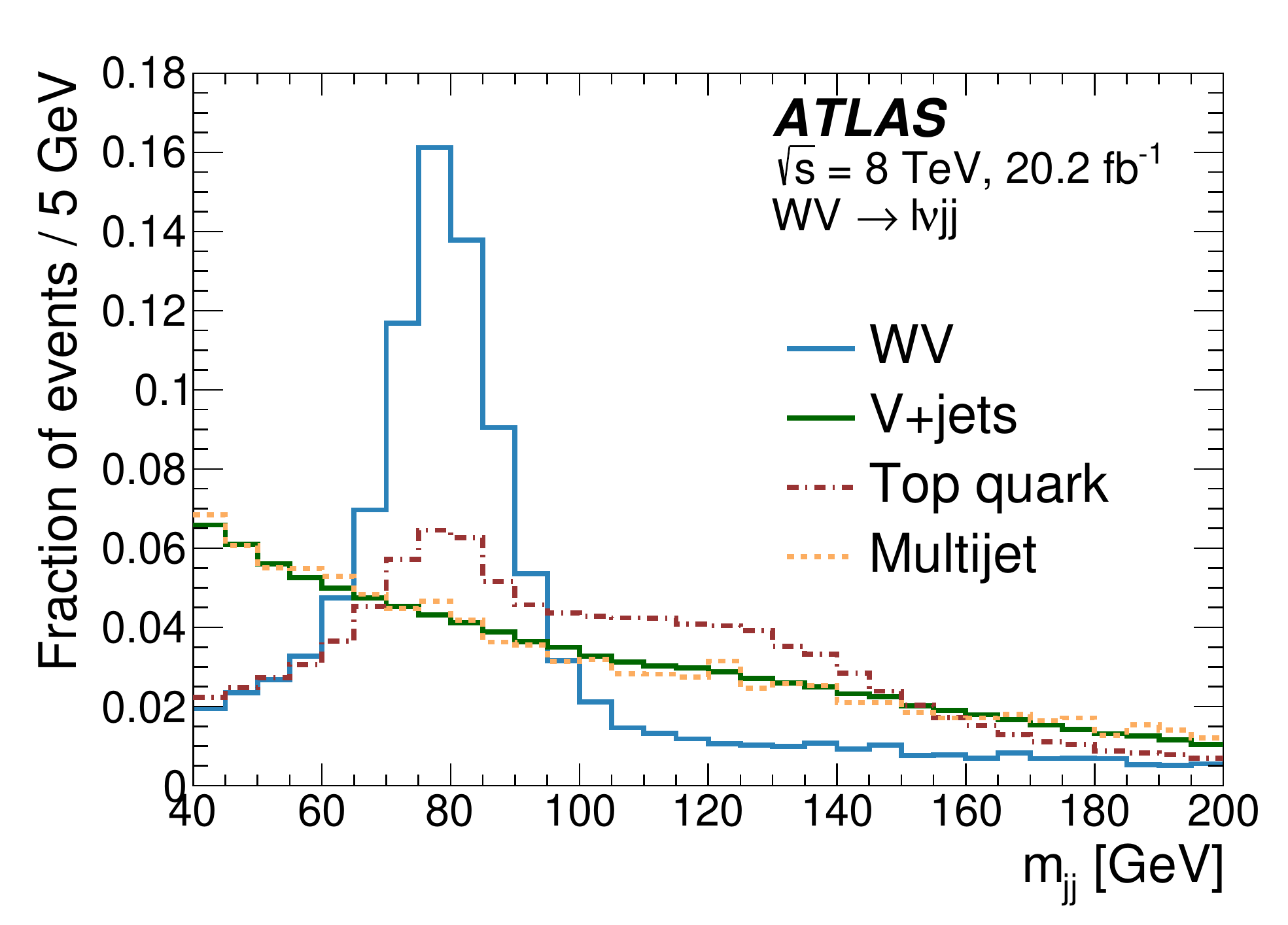}
}
\subfigure[]{
  \label{fig:templates_normalized_boosted}
 \includegraphics[width=0.47\textwidth,keepaspectratio]{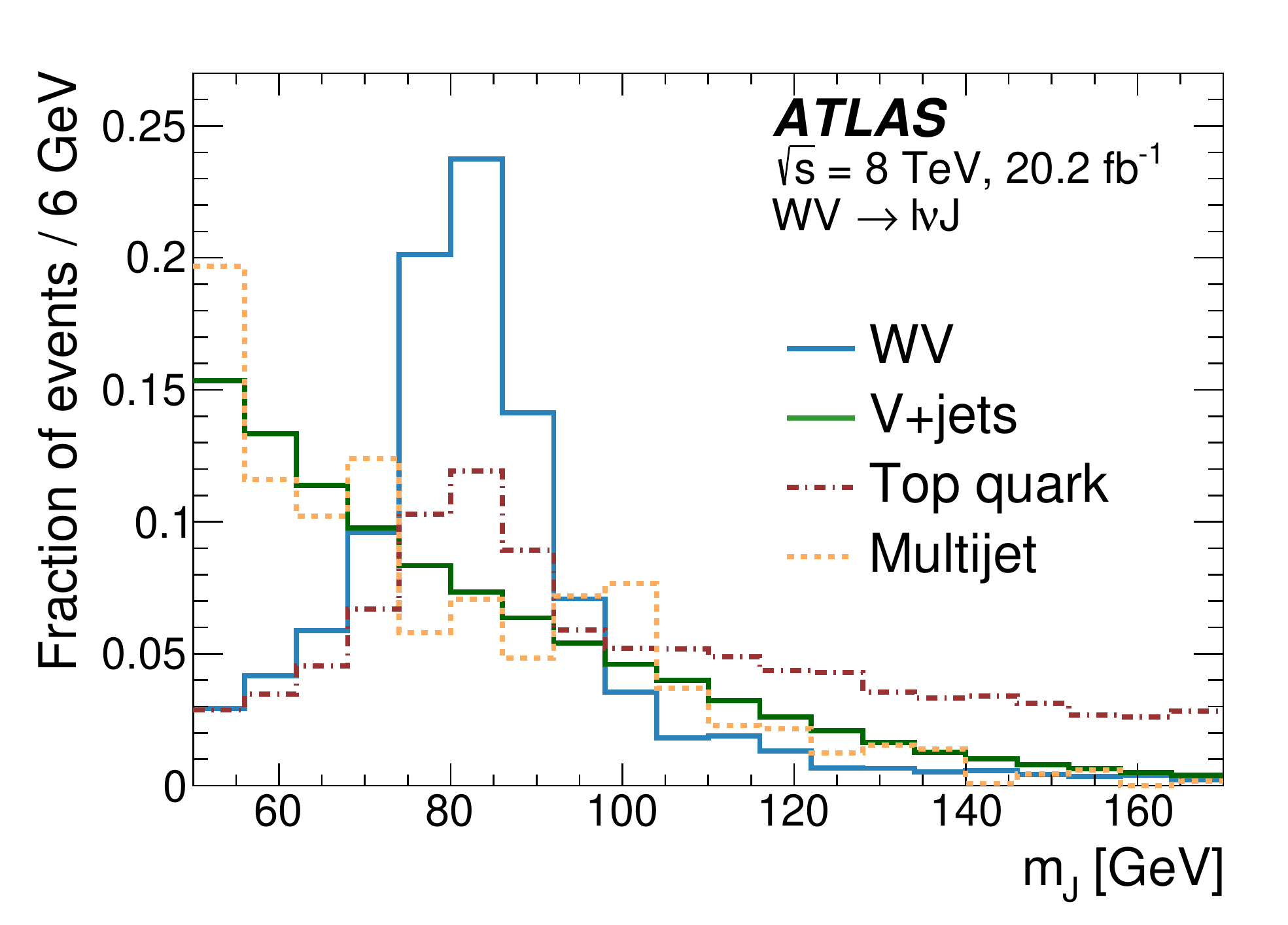}
}
\caption{
The shapes of (a) the predicted $\mjjless$ distributions in the \resolved\ signal region 
and (b) the predicted $\mJ$ distributions in the \boosted\ signal region, for the signal
(peaked near 80~\gev) and various background processes. The distributions are normalized to unity. 
}
\label{fig:templates_normalized}
\end{center}
\end{figure}

The expected signal and background yields in the \resolved\ signal region are 
given in Table~\ref{tab:yields}, and compared to the number of events observed in data.
The predictions for the $\mjjless$ distribution shapes of the signal
and backgrounds are shown in Figure~\ref{fig:templates_normalized_resolved}.

\begin{table}[htbp] 
\centering 
\begin{tabular}{l!!} 
\toprule
\midrule
& \multicolumn{1}{c}{\resolved} & \multicolumn{1}{c}{\boosted} \\
\midrule
\midrule
Signal\\
\midrule
$WW$  &  2860 , 110 & 542 , 61 \\ 
$WZ$  &  730 , 30 &  128 , 15 \\ 
Total Expected Signal  &  3590 , 140 & 670 , 75 \\ 
\midrule
Background\\
\midrule
$\Wjet$  &  136000 , 8600 &  10500 , 1300 \\ 
$\Zjet$  &  2750 , 340 &  245 , 32 \\ 
$\ttbar$  &  12980 , 520 &  1130 , 150 \\ 
Single top-quark  &  3620 , 150 &  249 , 35 \\ 
Multijet  &  3689 , 60 &  313 , 18 \\ 
 $ZZ$  & 14 , 1 & \multicolumn{1}{c}{-} \\ 
 Total Expected Background  &  159000 , 8600 &  12400 , 1500 \\ 
\midrule
Total SM Expected   &  162600 , 8700 &  13100 , 1600 \\ 
\midrule
Observed & \multicolumn{1}{c}{$164502$} & \multicolumn{1}{c}{$12999$} \\ 
\midrule
$S/B$ ($65~\gev<\mjjless<95~\gev$)   &  \multicolumn{1}{c}{$5.5\%$} & \multicolumn{1}{c}{$10.1\%$} \\
$S/\sqrt{B}$ ($65~\gev<\mjjless<95~\gev$)   &  \multicolumn{1}{c}{$11.1$} &  \multicolumn{1}{c}{$7.1$} \\
\bottomrule
\bottomrule
\end{tabular} 
\caption{ Expected number of signal and background events in the \resolved\ and \boosted\ signal regions,
prior to performing the $\mjjless$ and $\mJ$ fits.  The quoted uncertainties only include 
detector-related uncertainties and statistical uncertainties of the MC
samples and control regions.  The number of events observed in data is also shown.
The signal predictions only correspond to $\qq$-initiated $WV$ production.
}
\label{tab:yields} 
\end{table}

\subsection{\boosted\ channel}
\label{sec:backgrounds_boosted}

In the \boosted\ channel, the $\Wjet$, $\Zjet$, and top-quark
backgrounds are estimated using MC samples.  The  MC predictions for the 
two largest backgrounds ($\Wjet$ and top-quark production) are
corrected by scale factors obtained from dedicated control regions.

\begin{figure}[htbp]
\begin{center}
\subfigure[]{
  \label{fig:boostedTopCR_mJ}
 \includegraphics[width=0.47\textwidth,keepaspectratio]{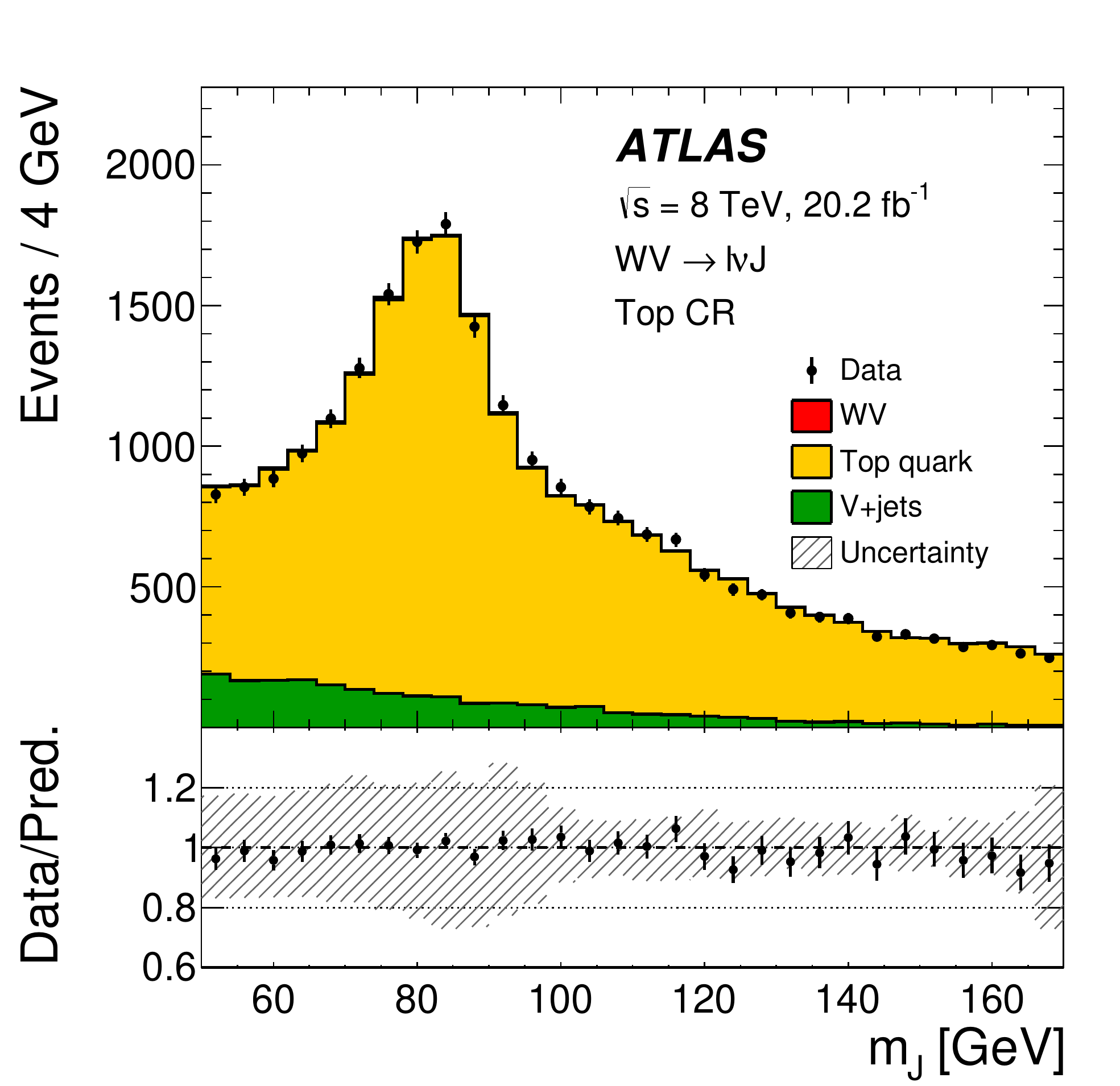}
}
\subfigure[]{
\label{fig:boostedWjetCR_pTJ}
  \includegraphics[width=0.47\linewidth,keepaspectratio]{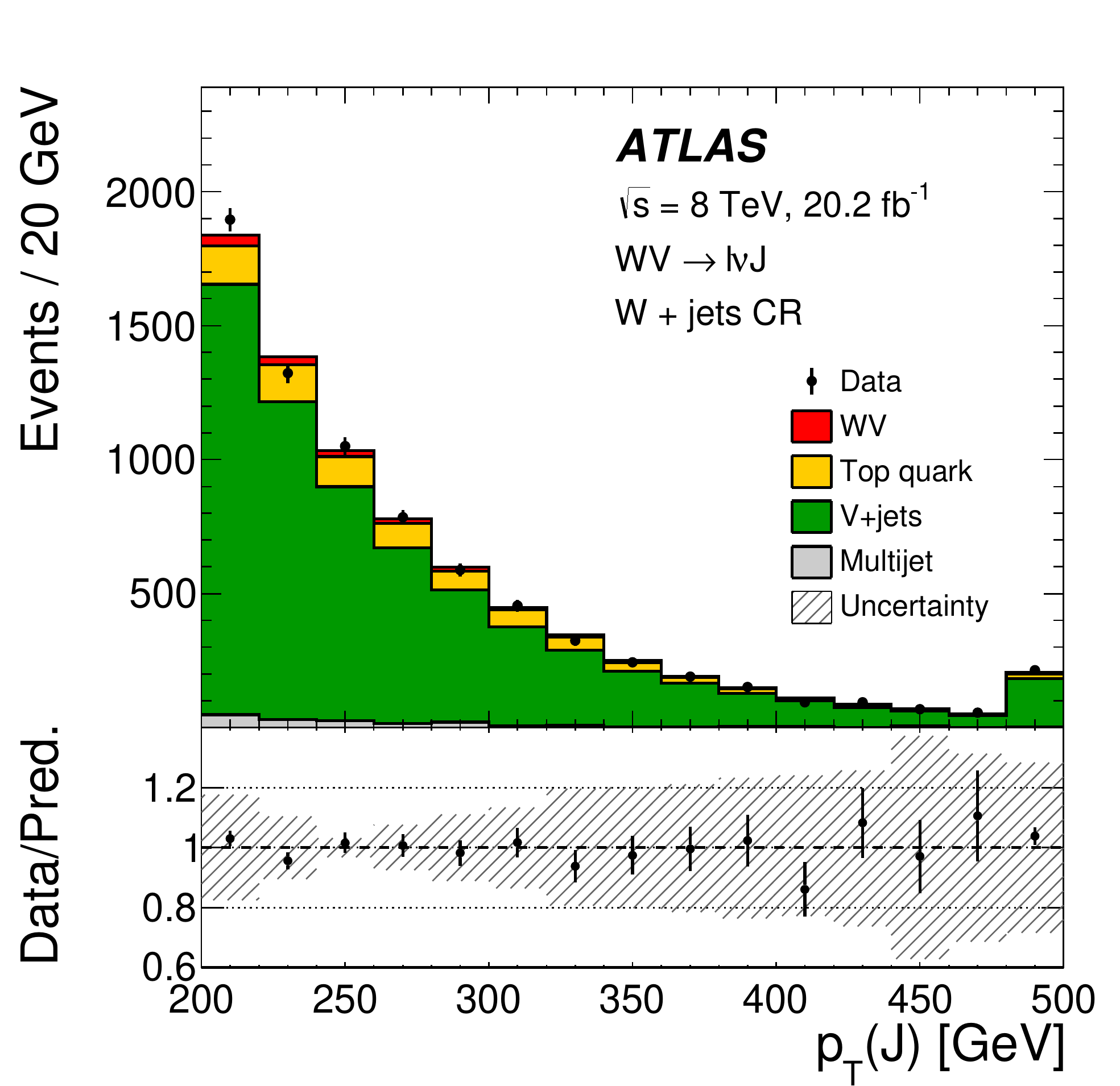}
}
\caption{ 
Comparison between data and prediction in the \boosted\ channel for (a) $\mJ$ in the top CR, and (b) $\pTJ$
in the $\Wjet$ CR.  A scale factor is applied to the top-quark background prediction in
the top CR and the $\Wjet$ CR, and a scale factor is applied to the $\Wjet$ background
prediction (which is part of the ``$\Vjet$'' histogram) 
in the $\Wjet$ CR.  
The hatched bands indicate the systematic uncertainty of the prediction.
For the $\Vjet$ component, only shape systematic uncertainties are included in the bands.
}
\label{fig:boostedCR}
\end{center} 
\end{figure}

The top-quark control region (top CR) is formed by events satisfying the signal
region selection, except that the presence of at least one small-$R$ $b$-tagged jet with 
$\pT>25~\gev$ and $\DR(\jet,\fatjet)>1.0$ is required instead of applying the nominal veto on small-R jets.  The jets are $b$-tagged using the MV1 
algorithm~\cite{ATLAS-CONF-2014-046,Aad:2015ydr}, using a working
point with a $b$-tagging efficiency of about 70\% and a gluon/light-quark jet rejection factor of over 100
in $\ttbar$ events.
About $90\%$ of the events in this top CR originate from top-quark backgrounds.
There is a deficit in data in the top CR relative to the MC prediction, which is attributed
to a mismodelling of the top-quark backgrounds. A global scale factor of $\topSFboost$ for the top-quark
backgrounds is obtained from this CR, after subtracting the prediction for non-top-quark backgrounds.
  The data in the top CR is shown in Figure~\ref{fig:boostedTopCR_mJ}, compared
to the SM prediction after application of the top-quark scale factor. 
This scale factor is applied to 
the top-quark background predictions in the signal region.
 
The control region for the $\Wjet$ background  ($\Wjet$ CR) is obtained 
by applying the standard signal region selection, but adding the requirement that
$\mJ<65~\gev$ or $\mJ>95~\gev$.
This additional $\mJ$ requirement removes almost all of the $WV$ signal events and 
also a large fraction of the top-quark events.
About $85\%$ of the events in this CR originate from $\Wjet$ backgrounds.
The top-quark background prediction in the $\Wjet$ CR is scaled by the top-quark scale factor obtained above.
A data deficit is observed in the $\Wjet$ CR relative to the prediction.  A global
scale factor of $\wjetSFboost$ is obtained for the $\Wjet$ background, after subtracting
the expected contributions from the other signal/background processes.
A comparison between the data and the prediction in the $\Wjet$ CR is shown in Figure~\ref{fig:boostedWjetCR_pTJ}, 
after application of the $\Wjet$
scale factor.  The $\Wjet$ scale factor is applied to the $\Wjet$ prediction in the signal region.

The method for estimating the multijet background is similar to that used 
in the \resolved\ channel.  As in the \resolved\ channel, a multijet
CR is defined by requiring an ``anti-identified'' lepton candidate.
The shapes of the kinematic distributions are estimated from this CR using the same
method as in the \resolved\ channel. The non-multijet background contributions to the CR are about 6\%
of the total.  The multijet event yield
is estimated from a fit to the $\met$ distribution, as in the \resolved\ channel, but
the only requirement that is removed for the definiton of the extended signal region/multijet CR 
is the $\met>50~\gev$ requirement.
The multijet background is found to be negligible for the muon channel, so only the
contribution in the electron channel is considered for the final results.

The numbers of expected and observed events in the \boosted\ signal region 
are summarized in Table~\ref{tab:yields}.  The previously mentioned top-quark and $\Wjet$
scale factors are applied to the predictions.  The contribution from $ZZ$ events is 
expected to be very small in the \boosted\ channel, so it is neglected.
 The nominal predictions for the $\mJ$ distribution shapes of the signal
and backgrounds are shown in Figure~\ref{fig:templates_normalized_boosted}.

%-------------------------------------------------------------------------------

%-------------------------------------------------------------------------------
\section{Cross-section extraction}
\label{sec:xsecMethod}
\label{sec:fiducialXsec}
The fiducial cross-section $\sigmafid$ for $\WVlnuqq$ production is measured 
independently for the \resolved\ and \boosted\ phase spaces, in both cases
using the formula:
\begin{eqnarray}
\label{eq:simple1}
\sigmafid  & = & \frac{\Nmeas}{\Lumi \cdot \Dfid} \,, \nonumber
\end{eqnarray}
where $\Nmeas$ is the measured signal yield, $\Lumi$ is the integrated luminosity,
and $\Dfid$ is a factor that corrects for experimental acceptance and efficiencies.
Since this analysis measures $\Nmeas$ as the sum of the $WW$ and $WZ$ processes,
which can each have different acceptances and efficiencies, $\Dfid$ is given by:
\begin{eqnarray}
\label{eq:simple2}
\Dfid  =   \fWWfid \cdot \CWW + (1-\fWWfid) \cdot \CWZ\,, \nonumber
\end{eqnarray}
where the $\CWV$ are reconstruction correction factors 
and the variable $\fWWfid$ is the predicted ratio of the $WW$ fiducial cross-section 
to the $WW+WZ$ fiducial cross-section.  
The  $\CWV$ and $\fWWfid$  values are estimated from MC simulation. 
The $\CWV$ factors are defined as the predicted 
number of $WV$ signal events passing the reconstruction-level event selection 
divided by the number of $WV$ events in the fiducial phase space defined with
generator-level particles.  The $\CWV$ factors account for reconstruction inefficiencies,
 resolution effects, and for contributions to the signal region from $WV$ events that do not 
decay to $\lnuqq$ (such as $\WVtaunuqq$ or $\WWlnulnu$); the latter are 
included in the $\CWV$ numerator and not in the denominator.
 The cross-section $\sigmafid$ is measured for the sum of
the electron and muon channels, so $\Dfid$ is computed as a weighted 
average over the
electron and muon channels. The fiducial cross-section measurement therefore 
assumes that the signal MC simulation correctly predicts the ratio of $WW$ to $WZ$
and of electrons to muons.
The value of $\Dfid$ is $\DfidValResolved$ in the \resolved\ channel and $\DfidValBoosted$
in the \boosted\ channel, including systematic uncertainties (see Section~\ref{sec:systematics}).

The fiducial phase spaces for the \resolved\ and \boosted\ channels 
are defined in Sections~\ref{sec:fiducialResolvedDef} 
and \ref{sec:fiducialBoostedDef}, respectively. These fiducial phase spaces partially overlap. 
In order to cope with the small signal-to-background ratios in this analysis (5--10$\%$),
the cross-section $\sigmafid$ is extracted using a binned ML fit to the $\mjjless$ distribution (in the
\resolved\ analysis) or the $\mJ$ distribution (in the \boosted\ analysis).  
The ML fits are performed on the sum of the electron and muon channels. It was cross-checked that
the electron and muon channels are compatible, in both the \resolved\ and \boosted\ channels.

In the ML fits, the value of $\sigmafid$  and the $\Vjet$ background yield are both free to vary without constraint.
Systematic uncertainties in the signal and backgrounds are incorporated in the fit by
including nuisance parameters that are allowed to vary within prior constraints. The
nuisance parameters allow the luminosity, $\Dfid$, the non-$\Vjet$ background yields,
and the $\mjjless$ and $\mJ$ shapes of the signal and background
distributions to vary within their systematic uncertainties.
The correlations between the uncertainty in $\Dfid$ and the uncertainty in the 
signal $\mjjless/\mJ$ shapes are accounted for in the fit.  The sources of systematic uncertainty
and the methods to assess these uncertainties are described in detail in Section~\ref{sec:systematics}.

\subsection{\resolved\ fiducial phase space}
\label{sec:fiducialResolvedDef}

The \resolved\ fiducial phase space is defined to closely match the
experimental event selection.  The phase-space definition requires
a $WV$ pair with the bosons decaying as $\Vtoqq$ and  $\Wtolv$, where $\ell=e,\mu$.  Events
containing other kinds of $WV$ decay channels (such as $\WWlep$ events or 
$\WVtaunuqq$ with the $\tau$ decaying to $\ell+X$), are
not included in the fiducial phase-space definition. Such $WV$  events can still pass
the experimental event selection (where they are included in the signal
category), and they are accounted for in the $\Dfid$ definition.

Leptons selected in the fiducial region must have $\pT(\ell)>30~\gev$ and $|\eta(\ell)|<2.47$. 
The four-momentum of the lepton is modified by adding to it the
four-momenta of all the photons within $\DR=0.1$, excluding
photons produced by hadron decays.  Particle-level anti-$k_t$ $R=0.4$ jets are constructed
using as constituents all stable particles, excluding muons and neutrinos. Stable particles are 
defined as those having a mean lifetime of $\tau>30$~ps.
The particle-level jets must have $\pT>25~\gev$ and $|\eta|<2.5$.  Jets within
$\DR=0.2$ of a selected electron are rejected, and then leptons within $\DR=0.4$
of a remaining jet are rejected.  
The true $\met$ in the event is defined as the magnitude of the vector $\pT$ sum of all
the neutrinos. 

The event must have exactly one lepton and two $R=0.4$ jets matching the above definitions.
The remaining requirements for the fiducial phase space are summarized
in Table~\ref{tab:FVDefinition}, and are analogous to the experimental event selection,
but are defined using the lepton, $\met$, and particle-level jets described
in this section.

\begin{table}[tb]
\begin{center}
\begin{tabular}{l|c|c} 
\hline \hline
   & \resolved\  & \boosted\ \\
\hline
Lepton  &   \multicolumn{2}{c}{$N_{\ell}=1$ with $\pT>30~\gev$ and $|\eta|<2.47$} \\
        &   \multicolumn{2}{c}{ $\DR(\ell,\jet)>0.4$  }  \\
\hline
$\Wtolv$ & $\pT(\ell\nu)>100~\gev$ &  $-$  \\
         & $\MT>40~\gev$  & $-$ \\
\hline
\MET  & $\MET > 40~\gev$  & $\MET > 50~\gev$ \\
\hline
Jet   &  $\njet=2$ with $\pT>25~\gev$, $|\eta|<2.5$,  & $\nfatjet=1$ with $\pT>200~\gev$, $|\eta|<2.0$, \\
      &  $\DR(\jet,e)>0.2$  &   $\DR(\fatjet,\ell)>1.0$ \\
      &       & No small-$R$ jets with $\pT>25~\gev$, $|\eta|<4.5$, \\
      &                        &    $\DR(\jet,\fatjet)>1.0$,   $\DR(\jet,e)>0.2$ \\
      &   $40<\mjjless<200~\gev$    & $50<\mJ<170~\gev$ \\
      &   $\pTjj>100~\gev$ & $-$  \\
      &   $\Delta\eta(\jet,\jet)<1.5$  & $-$ \\
\hline
Global   &  $\Delta\phi(\jet_1,\MET)>0.8$  & $-$ \\
\hline\hline
\end{tabular}
\caption{
Summary of the fiducial phase-space definitions. All the specified selection criteria are applied at the particle level as specified in the text. The notations ``$\jet$'' and ``$\fatjet$'' refer to $R=0.4$ and $R=1.0$ jets, 
respectively, as explained in the text.
}
\label{tab:FVDefinition}
\end{center}
\end{table}

\subsection{\boosted\ fiducial phase space}
\label{sec:fiducialBoostedDef}
As in the \resolved\ channel, the fiducial phase-space definition requires
a $WV$ pair with  $\Vtoqq$ and  $\Wtolv$.  Leptons, \met, and particle-level $R=0.4$ jets
are defined in the same way as in the \resolved\ channel, 
except that two 
sets of leptons and small-$R$ jets are considered: 
 \textit{central} leptons (small-$R$ jets) are required to have $|\eta|<2.47$ ($|\eta|<2.5$),
and \textit{extended} leptons and small-$R$ jets are required to have $|\eta|<4.5$.
Particle-level large-$R$ jets are defined by applying 
the anti-$k_t$ algorithm with radius parameter $R=1.0$ to all stable particles, excluding muons and neutrinos.
No trimming is applied to these jets.  The
large-$R$ jets are required to have $\pT>200~\gev$ and  $|\eta|<2.0$.
Central (extended) small-$R$ jets that are within
$\DR=0.2$ of a central (extended) electron are rejected.
Then, central leptons are rejected if they are within $\DR=0.4$ 
of a remaining central small-$R$ jet.
Large-$R$ jets are rejected if they are within $\DR=1.0$ of any remaining central leptons.
Events are required to contain exactly one central lepton and one large-$R$ jet with the above definitions,
and events are  discarded if they contain any extended small-$R$ jets with $\DR(\jet,\fatjet)>1.0$.
The event must also have $\met>50~\gev$, and the large-$R$ jet must have a mass greater than $50~\gev$.
The fiducial phase-space definition is summarized 
in Table~\ref{tab:FVDefinition}.

%-------------------------------------------------------------------------------

%-------------------------------------------------------------------------------
\section{Systematic uncertainties}
\label{sec:systematics}
Systematic uncertainties in the measured $\sigmafid$ can be due
to uncertainties in $\Lumi$, $\Dfid$, and/or $\Nmeas$.  Uncertainties
in the measured $\Nmeas$ can in turn be due to uncertainties in 
the background yields or in the shapes of the kinematic
distributions ($\mjj$, $\mJ$) of the signal and backgrounds
(hereafter called ``shape uncertainties'').  The dominant systematic
uncertainties in the $\sigmafid$ measurement are those affecting
the measured $\Nmeas$.

A wide variety of detector-related experimental uncertainties are considered,
which affect $\Dfid$, the predicted background yields, and the signal and
background shapes.
The most important of these uncertainties are those related to the
jet reconstruction.  Uncertainties in the small-$R$ jet energy 
scale and resolution
are accounted for~\cite{PERF-2012-01,PERF-2011-04}.  
In the \boosted\ channel,  uncertainties in the large-$R$
jet energy and jet mass scales are also taken into account.  The
scale uncertaities of the large-$R$ jets are estimated using a double-ratio
method that compares calorimeter- and track-jets in data and MC
simulation~\cite{PERF-2015-03}.
The energy 
and mass resolution uncertainties of large-$R$ jets are estimated
by smearing the jet energies/masses so as to degrade the resolutions by 20\%; this
approach is based on prior studies of large-$R$ jets~\cite{STDM-2011-19,HIGG-2013-19}.
The systematic uncertainty due to the JVF requirement is also included~\cite{ATLAS-CONF-2013-083}.
In addition to the jet-related uncertainties, there are also 
systematic uncertainties in the electron and muon reconstruction
 (including triggering, object reconstruction, identification,   
and the energy scale and resolution)~\cite{Aaboud:2016vfy,PERF-2014-05,PERF-2013-05,ATLAS-CONF-2012-048,TRIG-2012-03}. 
The effects of the jet and lepton
uncertainties are propagated to the \met calculation, and an additional
systematic uncertainty in
the soft terms entering the \met calculation is also included~\cite{PERF-2014-04}.

In the cross-section fits, the $\Vjet$ yield
is taken to be a free parameter, while several uncertainties in the
modelling of its shape are accounted for (in addition to the
shape uncertainties from the previously mentioned detector effects).
Systematic uncertainties in the $\Vjet$ shape 
are estimated by varying  the MC event generator 
used (\SHERPA\ compared to \ALPGENPYTHIA). The differences between the predictions of the two generators are taken as additional systematic uncertainties. 
Additional uncertainties
in the $\Vjet$ shape are estimated 
by varying the renormalization and factorization scales by factors of $2$ and $0.5$, and 
by varying the scale used in \SHERPA\
for matching the matrix elements to the parton showers~\cite{Hoeche:2009rj} from its nominal 
value of $20~\gev$ to alternative values of $15~\gev$ and $30~\gev$.
In the \resolved\ channel, the uncertainty in the shapes of the $\Vjet$ predictions
due to the two kinematic reweighting functions (see Section~\ref{sec:backgrounds_resolved}) 
is estimated by including the full 
difference between applying and not applying each reweighting function
as additional systematic uncertainties.  In the \resolved\ channel,
an uncertainty of $10\%$ in
the $(\Wjet)/(\Zjet)$ cross-section ratio is also included; this 
uncertainty is ignored
in the \boosted\ channel as it has a negligible effect.

For the $\ttbar$ background, uncertainties due to the matrix-element event generator,
parton shower/hadronization model, and amount of initial- and final-state
radiation are all included.  
The theoretical uncertainties in the top-quark background
cross-sections are also taken into account.   In the \boosted\
channel, instead of using the theoretical cross-section uncertainty,
the top-quark background is assigned a normalization uncertainty of $14\%$
to account for the uncertainty in the data-driven scale factor.
Systematic uncertainties in 
the multijet background estimate are also included,
which affect both its normalization and its shape. These uncertainties
are derived from studies of variations of the data-driven estimate, such
as changing the control region definitions and varying the non-multijet
background subtraction.
The uncertainty in the multijet
yield amounts to $30\%$ ($100\%$) for the electron (muon) channel in the
\resolved\ channel.  In the \boosted\ channel, an  uncertainty of $50\%$
is assigned to the multijet yield in the electron channel, 
while the multijet background 
is neglected in
the muon channel.   A $30\%$
uncertainty is assigned to the $ZZ$ event yield in the \resolved\ channel, to account for uncertainties in 
the $ZZ$ cross-section and the extrapolation to the fiducial phase space.

Additionally, the uncertainty in the modelling of pile-up interactions is accounted for~\cite{Aaboud:2016rmg}. 
The uncertainty in the integrated luminosity is also included,
computed as described in Ref.~\cite{Aaboud:2016hhf}.
The statistical uncertainty of the MC samples is taken into account,
which affects each bin in the ML fits in an uncorrelated way.

Uncertainties in the signal shapes and in the $\Dfid$ parameter due to variations of the signal model
are computed by varying the renormalization and factorization scales by
factors of $2$ and $0.5$, and by comparing the 
nominal \MCatNLO\ signal samples to alternative samples
generated with \SHERPA\ and \POWHEG+\PYTHIA8. The effect on $\Dfid$ from
the uncertainties in the CT10 PDF set is also taken into account;
the PDF uncertainty has a negligible impact on the signal shapes.  

The measured $\sigmafid$ values are compared to theoretical predictions from \MCatNLO. 
The uncertainty in the theoretical $\sigmafid$ prediction is calculated including the
uncertainties due to renormalization and factorization scales.
 Since the fiducial phase spaces contain a veto on additional jets, the 
Stewart--Tackmann procedure~\cite{Stewart:2011cf} is used to estimate the scale uncertainties. These
uncertainties are also propagated to the theoretical $\fWWfid$ value which enters into
the $\Dfid$ calculation, although the effect of this on the measured $\sigmafid$ is 
very small ($\sim 0.1\%$).  PDF-induced uncertainties in the theoretical prediction are
also taken into account.

%-------------------------------------------------------------------------------

%-------------------------------------------------------------------------------
\section{Cross-section results}
\label{sec:xsecResult}
\begin{figure}[htbp]
\begin{center}
 \includegraphics[width=0.47\textwidth,keepaspectratio]{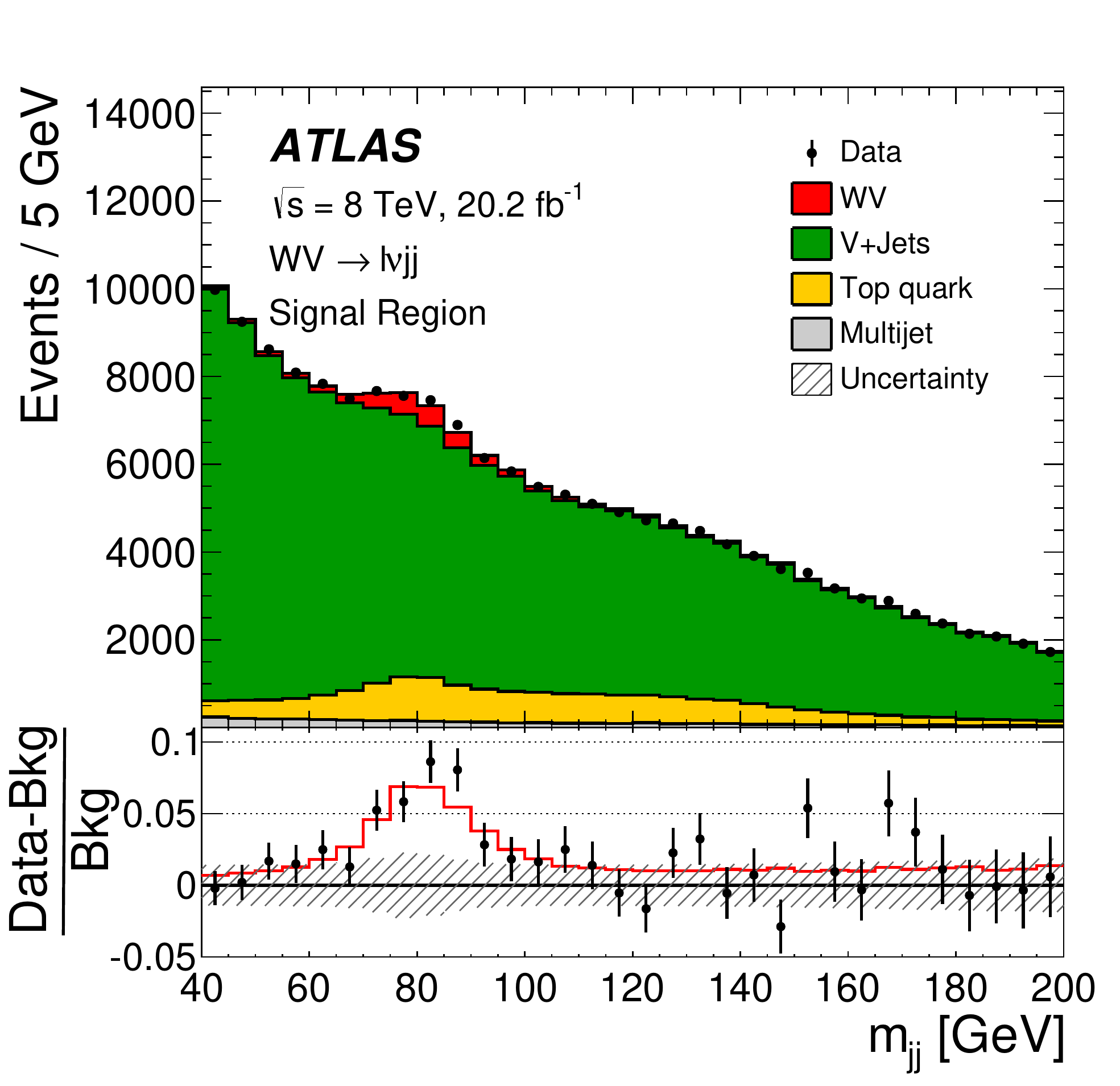}
\caption{
The observed $\mjj$ distribution in the \resolved\ signal region, overlaid with 
the post-fit background and \signal\ estimates.
The hatched band indicates the total uncertainty of the fit result.
}
\label{fig:resolved_postfit_mjj}
\end{center}
\end{figure}

The result of the ML fit to the $\mjjless$ distribution
 for the \resolved\ channel is shown 
in Figure~\ref{fig:resolved_postfit_mjj}. The fit is performed on the sum of events in the
electron and muon channels.
The observed significance is $\resultSignifResolvedObs$,
including statistical and systematic uncertainties,\footnote{The significance is calculated
based on the profile-likelihood ratio of the background-only and signal-and-background
hypotheses.  This ratio is converted to a significance using the asymptotic approximation~\cite{Cowan:2010js}.}
while the expected significance, calculated
using the Asimov data set~\cite{Cowan:2010js}, is $\resultSignifResolvedExp$.
The fitted $\Vjet$ background normalization is $1.02\pm 0.01$ times its pre-fit value, while the fitted
top-quark background normalization is $0.96\pm 0.10$ times its pre-fit value.

 The fiducial cross-section for the signal process
is extracted from the fit as described in Section~\ref{sec:xsecMethod},
and the result is 
\begin{eqnarray}
\sigmafidRobs  & = &  \resultSigmafidResolvedObs \,. \nonumber
\end{eqnarray}
The impacts of the various systematic uncertainties on the cross-section measurement are shown in Table~\ref{tab:systBreakDownResolved}.
The measurement can be compared to the theoretical prediction of
\begin{eqnarray}
\sigmafidRtheory  & = & \resultSigmafidResolvedTheory \,. \nonumber
\end{eqnarray}
The theoretical prediction is obtained using \MCatNLO\ for
the $\qqtoWV$ prediction.  The $\ggtoWW$ prediction is
also included, and is calculated using the 
total NLO $\ggtoWW$ cross-section prediction~\cite{Caola:2109129}
multiplied by the $\qqtoWW$ acceptance from \MCatNLO.
The $\ggtoWW$ contribution increases the fiducial cross-section 
prediction by $4\%$ in both the \resolved\ and \boosted\ channels.
Given the relatively small $\ggtoWW$ contribution, the possible
differences in acceptance between the 
$\ggtoWW$ and $\qqtoWW$ processes are neglected.
The uncertainty in the \MCatNLO\ prediction is described 
in Section~\ref{sec:systematics}.

\begin{table}[htbp] 
\centering 
\begin{tabular}{lr} 
\toprule
\midrule
Source of uncertainty  & Relative uncertainty for \sigmafid \\
\midrule\midrule
Top-quark background modelling & 13\% \\
Signal modelling & 12\% \\
$\Vjet$ modelling & 4\% \\
Multijet background modelling & 1\% \\
\midrule
Small-$R$ jet energy/resolution &  9\%  \\
Other experimental (leptons, pile-up)  & 4\% \\  
Luminosity &   2\%  \\
\midrule
MC statistics & 9\% \\
Data statistics & 14\% \\
\bottomrule
\bottomrule
\end{tabular} 
\caption{ Breakdown of the uncertainties in the measured fiducial cross-section in the \resolved\ channel.
Uncertainties smaller than 1\% are omitted from the table.}
\label{tab:systBreakDownResolved}
\end{table}

\begin{figure}[htbp]
\begin{center}
 \includegraphics[width=0.47\textwidth,keepaspectratio]{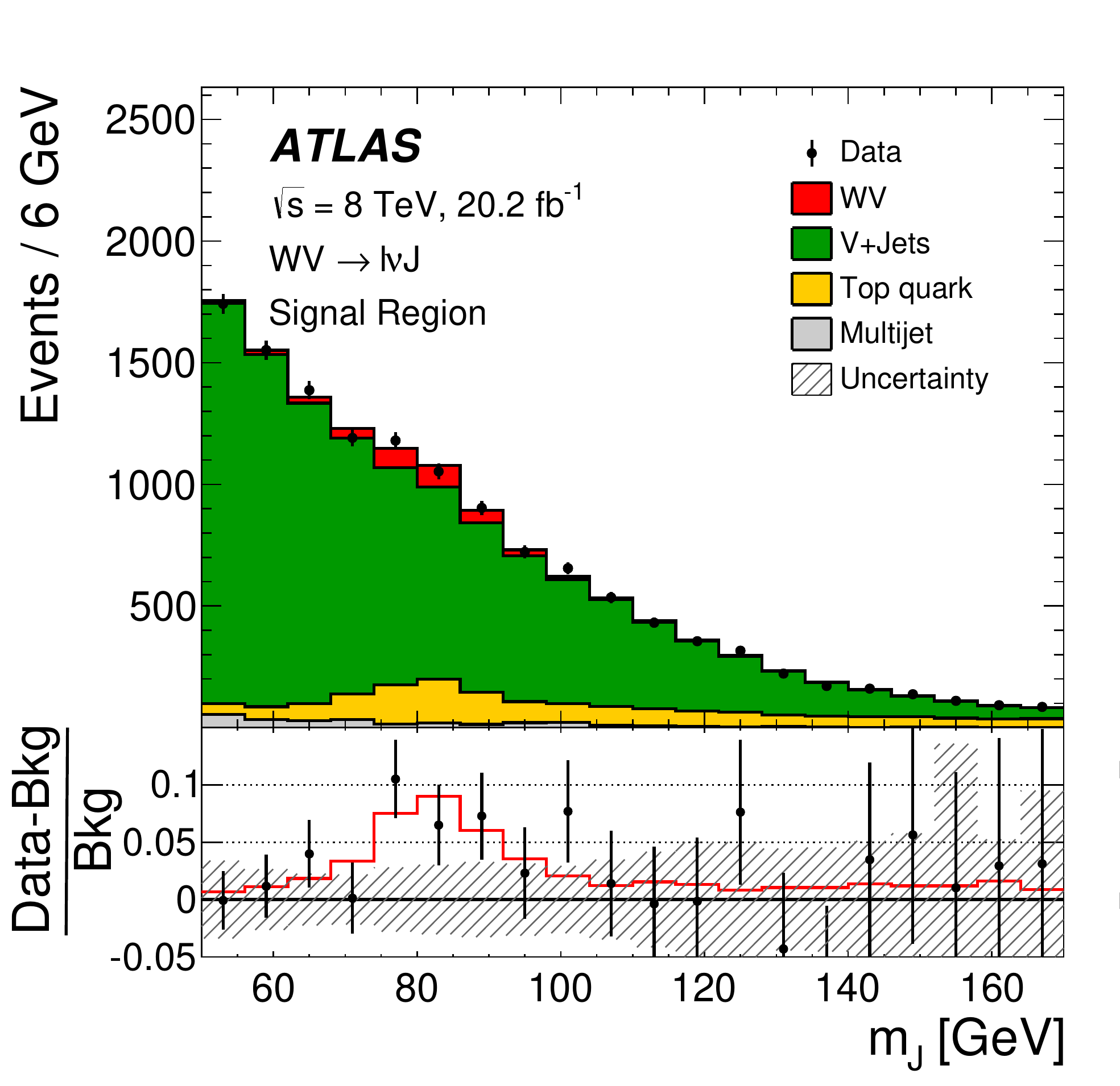}
\caption{
The observed $\mJ$ distribution in the \boosted\ signal region, overlaid with 
the post-fit background and \signal\ estimates.
The hatched band indicates the total uncertainty of the fit result.
}
\label{fig:boosted_postfit_mJ}
\end{center}
\end{figure}

The result of the $\mJ$ fit for the \boosted\ channel
is shown in Figure~\ref{fig:boosted_postfit_mJ}.  
Although the signal-to-background
ratio is better in this case than in the \resolved\ channel,
the total number of signal events is much smaller.  The
observed significance of the result is $\resultSignifBoostedObs$ (including
statistical and systematic uncertainties), compared
to an expected significance of $\resultSignifBoostedExp$.  
The fitted $\Vjet$  (top-quark) background normalization is $1.01\pm 0.04$   ($1.06\pm 0.20$) 
times its pre-fit value.

The extracted fiducial cross-section for the signal process is
\begin{eqnarray}
\label{eq:xsecBoostedObs}
\sigmafidBobs  & = &  \resultSigmafidBoostedObs \,, \nonumber
\end{eqnarray}
which is compatible with the theoretical prediction of
\begin{eqnarray}
\label{eq:xsecBoostedTheory}
\sigmafidBtheory  & = & \resultSigmafidBoostedTheory \,. \nonumber
\end{eqnarray}
The breakdown of the uncertainties contributing to the fiducial cross-section measurement
is shown in Table~\ref{tab:systBreakDownBoosted}.

\begin{table}[htbp] 
\centering 
\begin{tabular}{lr} 
\toprule
\midrule
Source of uncertainty  & Relative uncertainty for \sigmafid \\
\midrule\midrule
$\Vjet$ modelling & 60\% \\
Top-quark background modelling & 32\% \\
Signal modelling & 15\% \\
Multijet background modelling & 13\% \\
\midrule
Large-$R$ jet energy/resolution &  45\%  \\
Small-$R$ jet energy/resolution &  16\%  \\
Other experimental (leptons, pile-up)  & 3\% \\  
Luminosity &   2\%  \\
\midrule
MC statistics & 19\% \\
Data statistics & 33\% \\
\bottomrule
\bottomrule
\end{tabular} 
\caption{ Breakdown of the uncertainties in the measured fiducial cross-section in the \boosted\ channel.
Uncertainties smaller than 1\% are omitted from the table.}
\label{tab:systBreakDownBoosted}
\end{table}

\begin{figure}[htbp]
\begin{center}
 \includegraphics[width=0.6\textwidth,keepaspectratio]{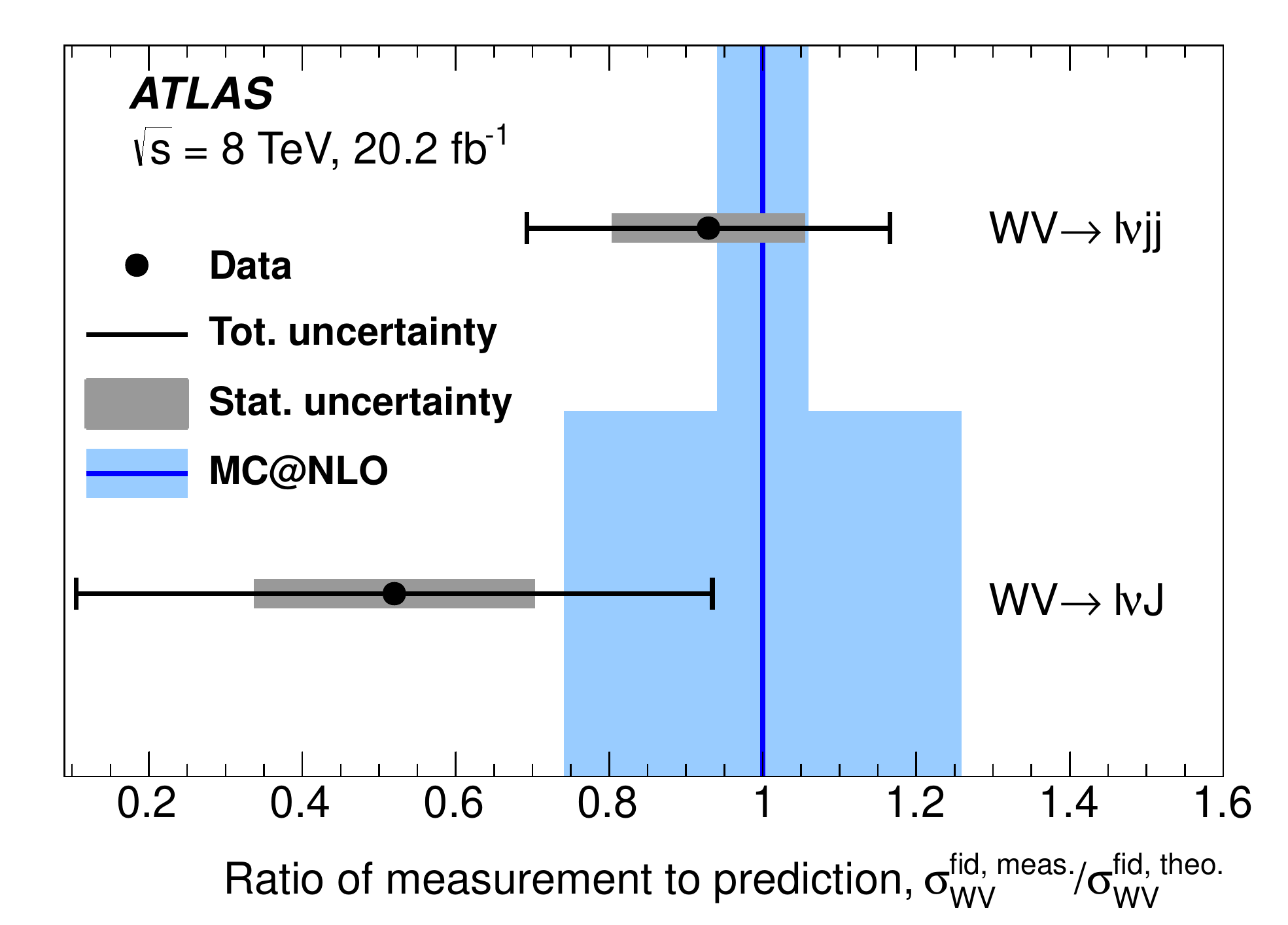}
\caption{ 
The ratios of the measured fiducial cross-sections to the cross-sections
predicted by \MCatNLO, for  the \resolved\ and \boosted\ 
phase spaces.  The \resolved\ and \boosted\ phase spaces partially overlap.
}
\label{fig:fidxsec_meas_theory_ratio}
\end{center}
\end{figure}

The cross-section measurements are summarized  
in Figure~\ref{fig:fidxsec_meas_theory_ratio}.
As mentioned in Section~\ref{sec:fiducialXsec}, the two
cross-section measurements are performed in partially overlapping
phase spaces.  The uncertainty in the theory prediction is significantly
larger in the \boosted\ channel than in the \resolved\ channel.
The theoretical uncertainty in the \boosted\ channel is dominated
by the scale uncertainties, which are particularly large because of
the aggressive jet veto in this channel (only about 30\% of signal
MC events pass the jet veto in the \boosted\ channel, 
compared to about 80\% in the \resolved\ channel).

%-------------------------------------------------------------------------------

%-------------------------------------------------------------------------------
\section{Constraints on anomalous gauge couplings}
\label{sec:aTGC}
In many extensions of the SM, diboson production can be modified, such
as through new resonances that couple to bosons.
If the scale of new physics is sufficiently high, new resonances may not
be visible in the current data; however,  diboson
production could still be affected below the new-physics scale,
in the form of modified couplings.  One common framework for parameterizing
new physics in diboson production is an effective Lagrangian~\cite{Hagiwara:1986vm}
of the form:
\begin{eqnarray}
   {\cal L}^{WWX} &  \propto & \Big[ (1+\dgv) (W^+_{\mu\nu}W^{-\mu} - W^{+\mu}W^-_{\mu\nu})X^\nu   \nonumber \\
                  &          & + (1+\dkv) W^+_\mu W^-_\nu X^{\mu\nu}  + \frac{\lambdav}{m_W^2} W^{+\nu}_\mu W_\nu^{-\rho}X^\mu_\rho  \Big]\,, \nonumber
\end{eqnarray}
where $X=Z$ or $\gamma$, $W^{\pm}_{\mu\nu}=\partial_\mu W^{\pm}_\nu - \partial_\nu W^{\pm}_\mu$,
and $X_{\mu\nu}=\partial_\mu X_\nu - \partial_\nu X_\mu$.  The six parameters
$\lambdav$, $\dkv$, and $\dgv$ (hereafter called ``aTGC parameters'')
are all zero in the SM.  The parameter $\dgg$ is zero because of EM gauge
invariance, leaving five free aTGC parameters, which describe deviations
 of the triple gauge-boson couplings from their SM predictions.
It is common to apply 
the so-called {\em LEP constraint}~\cite{Gounaris:1996rz},
which imposes $SU(2)\times U(1)$ gauge invariance, and which reduces the number of 
independent aTGC parameters to three, by introducing the following constraints:
$\lambdag=\lambdaz$ and $\dgz=\dkz + \dkg \tan^2\theta_W$, where $\theta_W$ is the
weak mixing angle.
Since aTGC parameters lead to violation of unitarity
at high energies, form factors are often applied to them in order
to ensure unitarity:
\begin{eqnarray}
   \alpha & \to & \frac{\alpha}{\left(1 + \frac{\shat}{\lambdaff^2}\right)^2}\,,  \nonumber
\end{eqnarray}
where $\alpha$ is one of the aTGC parameters, $\shat$ is the square of the diboson
invariant mass, and $\lambdaff$ is the form factor's energy scale.

An alternative framework for describing modifications of diboson production 
is an effective field theory (EFT)~\cite{HISZ:1993,Degrande:2012wf}
that is assumed to be valid below an energy scale $\lambdanp$, 
and which introduces three CP-conserving dimension-six operators:
\begin{eqnarray}
   {\cal O}_{W} & = & (D_\mu \Phi)^{\dagger} W^{\mu \nu} (D_\nu \Phi)\,, \nonumber \\
   {\cal O}_{B} & = & (D_\mu \Phi)^{\dagger} B^{\mu \nu} (D_\nu \Phi)\,, \nonumber \\ 
  {\cal O}_{WWW} & = & {\normalfont Tr}[ W_{\mu \nu} W^{\nu \rho} W^{\mu}_{\rho}] \,. \nonumber
\end{eqnarray}
Here, $\Phi$ is the Higgs doublet field, $D_\mu$ is the covariant derivative,
and $ W^{\mu \nu}$ and $B^{\mu \nu}$ are the field strength tensors of the $W$ and $B$
gauge boson fields.
The coefficients of these operators (EFT parameters), $\cw$, $\cb$, and $\cwww$, are zero in the SM and can be
related to the LEP-constraint aTGC parameters as follows:
\begin{eqnarray}
 \cwf & = & \frac{2}{m_Z^2}\dgz\,, \nonumber \\
\cbf & = & \frac{2}{m_W^2}\dkg  - \frac{2}{m_Z^2}\dgz\,, \nonumber \\ 
\cwwwf & = &  \frac{2}{3 g^2 m_W^2} \lambda\,. \nonumber
\end{eqnarray}
This relation only holds if no form factor is applied to the aTGCs.
The effect of aTGC/EFT parameters on the $H\to WW$ process
is neglected.

The aTGC and EFT parameters both tend to increase the diboson cross-section at
high $\pT(V)$ and high invariant mass of the diboson system. Both the \resolved\
channel and the \boosted\ channel can be used to search for these BSM enhancements.
The \boosted\ channel, although currently less sensitive as a SM $WV$ measurement,
is expected to provide a higher sensitivity to the aTGC/EFT models, because of the
better efficiency at high $\pT(V)$.  On the other hand, the \resolved\ channel, where
the SM $WV$ measurement is clearly established, is useful as a complementary
search channel that probes a different energy range.

In this analysis, the new-physics search uses 
signal regions with exactly the same 
event selection as the cross-section measurements,
except that the $\mjj$ requirement is tightened to 
 $65~\gev<\mjjless<95~\gev$ in the \resolved\ channel and the $\mJ$ requirement
is tightened to  $65~\gev<\mJ<95~\gev$ in the \boosted\ channel. These tighter
requirements lead to an increase in the signal-to-background ratio.
In the \resolved\ channel, events which fail the $\mjj$ requirement
(i.e. $40~\gev<\mjj<65~\gev$ or $95~\gev<\mjj<200~\gev$)
are put into a {\em sideband} control region.
The $ZZ$ background is neglected in the new-physics search, due to its
very small expected contribution.

\begin{figure}[htbp]
\begin{center}
\subfigure[]{
 \includegraphics[width=0.47\textwidth,keepaspectratio]{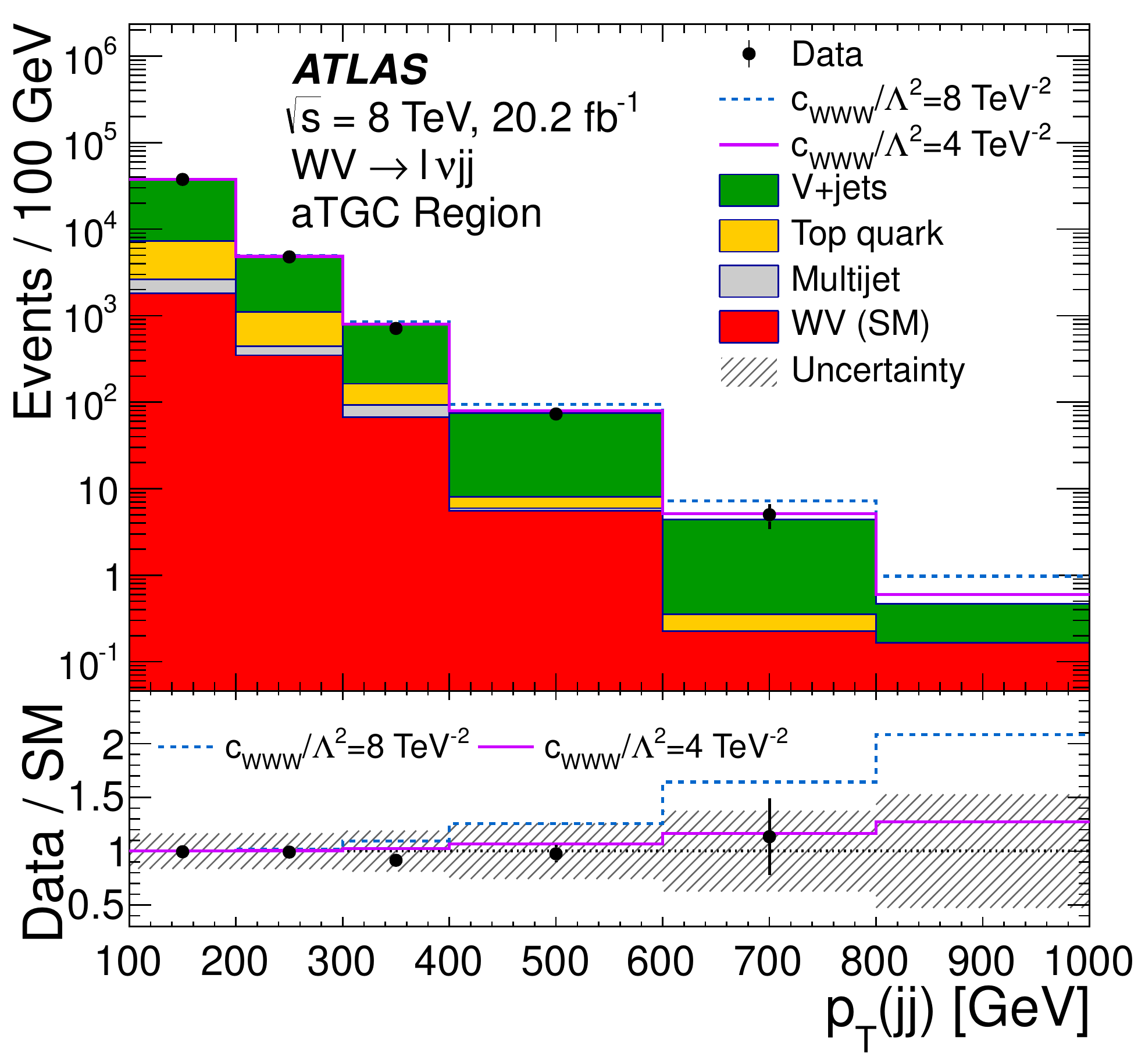}
}
\subfigure[]{
 \includegraphics[width=0.47\textwidth,keepaspectratio]{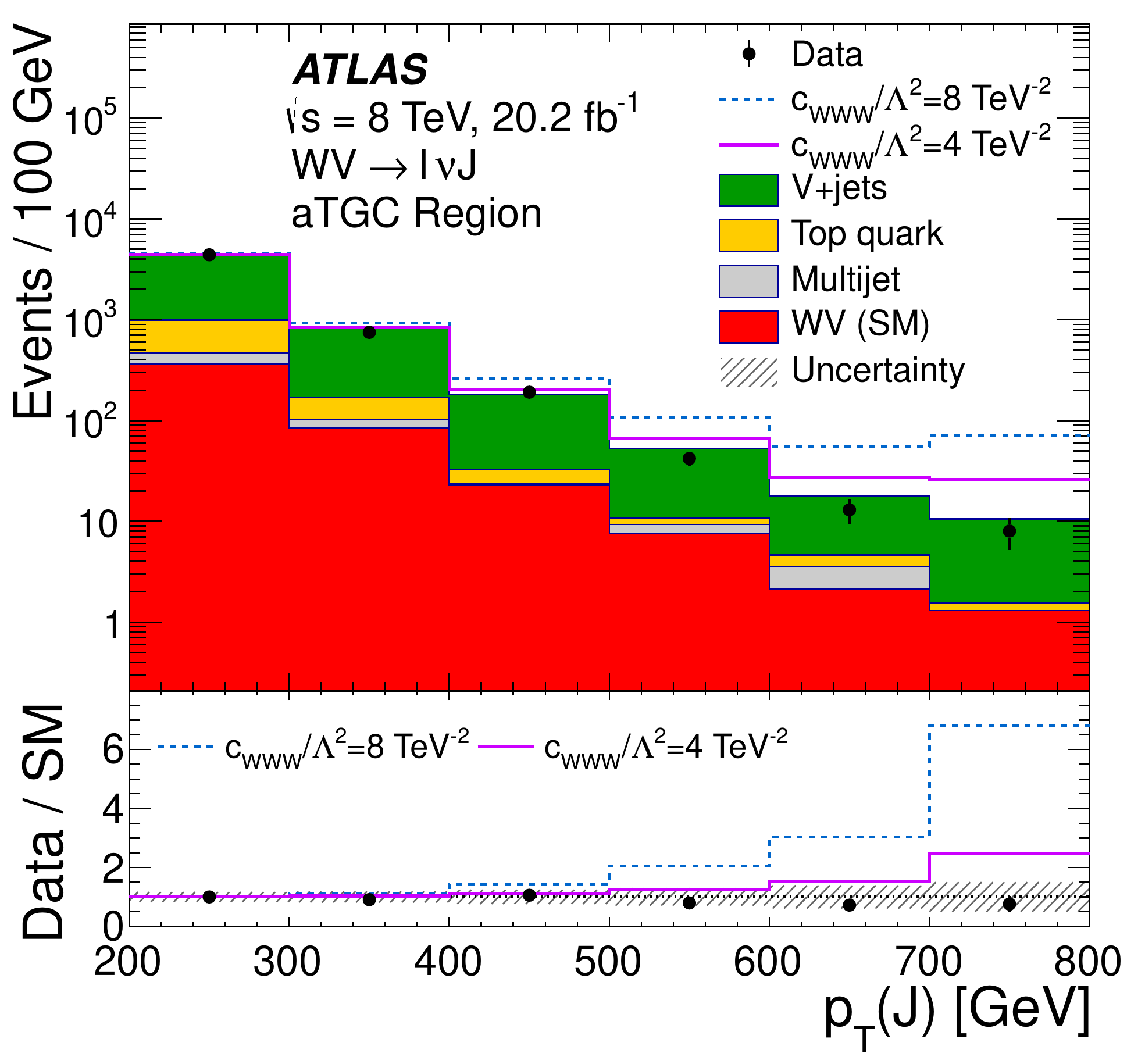}
}
\caption{
 The observed (a) $\pTjj$ distribution in the \resolved\ 
\atgcSR, and (b) $\pTJ$ distribution in the \boosted\ 
\atgcSR, overlaid with the background and \signal\ prediction.
The expected BSM enhancements due to anomalous values of the EFT parameter $\cwww$ 
are also shown, for $\cwww=4~\tevminSq$ and $\cwww=8~\tevminSq$. 
The hatched bands indicate the systematic uncertainty in the SM prediction.
The histograms are displayed with the binning that is used for the computation
of the confidence intervals for the aTGC and EFT parameters.
The last bin includes overflow.
}
\label{fig:aTGCfitDistObservedBoosted}
\end{center}
\end{figure}

The search makes use of the $\pTjj$ (\resolved\ channel)
or  $\pT(\fatjet)$ (\boosted\ channel) distribution. Hereafter, $\pTVhad$
is used to refer to both $\pTjj$ and $\pT(\fatjet)$.
 The $\pTVhad$ distributions  of the events in the signal regions are shown 
in Figure~\ref{fig:aTGCfitDistObservedBoosted}.  
This figure also shows the expected
enhancement at high $\pTVhad$ in the presence of different EFT parameter values.
As can be seen from the figure, no significant deviation from the SM prediction
is observed; therefore, $95\%$ confidence 
intervals are computed for the aTGC and EFT parameters.

The confidence intervals are computed from 
binned ML fits to the $\pTVhad$ distributions.  
The intervals are calculated using a frequentist Feldman--Cousins approach~\cite{ref:FeldmanCousins}.
In  the \resolved\ channel, simultaneous fits to the $\pTVhad$ distributions in
the signal region and sideband CR are used, while in the \boosted\ channel, only
the  $\pTVhad$ distribution in the signal region is used.
Since the \boosted\ and \resolved\ selections overlap, the confidence intervals are calculated separately
for the \boosted\ and \resolved\ selections.  
In the fits, the SM $WV$ and background predictions are allowed to vary within their uncertainties.
The measured cross sections of Section~\ref{sec:xsecResult} are consistent with theoretical SM $WV$ predictions, but have large associated uncertainties; for this reason the theoretical prediction is used here. 
The systematic uncertainties in the normalizations and $\pTVhad$ shapes of the signal and
backgrounds are accounted for through nuisance parameters.  
The systematic uncertainties that have the largest impact on the results are the jet-related uncertainties
(in both channels) and the uncertainty from the limited size of the MC samples (in the \resolved\ channel).

The observed 95\% confidence intervals for the aTGC parameters are
shown in Table~\ref{tab:limits_aTGCnoconstraint}, without applying the LEP constraint.
The confidence intervals for a given aTGC parameter are computed while fixing the other
aTGC parameters to zero.
The confidence intervals are shown  separately for the 
\resolved\ and \boosted\ selections, and the expected confidence intervals under the 
SM hypothesis are also shown for comparison.  
Confidence intervals for the aTGC parameters are shown for  $\lambdaff=\lffval~\tev$ and
for the case of no form factor (equivalent to $\lambdaff=\infty$).  The value of
 $\lambdaff=\lffval~\tev$ is chosen in order to ensure unitarity over the range
of aTGC parameter values to which this analysis is sensitive~\cite{Aihara:1995iq}.

The \boosted\ selection has significantly better sensitivity
to the aTGC parameters.  No combination of the \resolved\ and \boosted\ constraints
is performed, since it is expected that the \boosted\ channel would dominate the
combination.
The sensitivity to the aTGC parameters in the \boosted\ channel mainly comes from the $\pTVhad>600~\gev$ bins, 
whereas the sensitivity in the \resolved\ channel mainly comes from
the 300--600$~\gev$ bins.  Since the \resolved\ channel probes a lower $\pTVhad$
range, its sensitivity is less degraded by the form factors (which have a larger
effect at higher $\pT$) than the \boosted\ channel.

In addition, the observed  and expected confidence intervals for the aTGC parameters
in the LEP-constraint scenario are given in Table~\ref{tab:limits_aTGClep} for
$\lambdaff=\infty$.

\begin{table}
  \begin{center}
\begin{tabular}{lccccc}
  \hline \hline
Form factor   & Parameter   &  Observed  & Expected  &  Observed  & Expected   \\
           &   &   \multicolumn{2}{c}{ \resolved } &  \multicolumn{2}{c}{\boosted} \\
\hline
  & \dgz & [ $-0.039, 0.059$] & [ $-0.050, 0.066$] & [ $-0.033, 0.036$] & [ $-0.039, 0.042$]\\
   & \dkz & [ $-0.045, 0.063$] & [ $-0.060, 0.076$] & [ $-0.028, 0.030$] & [ $-0.033, 0.035$]\\
    $\lambdaff=\infty$ & \lambdaz & [ $-0.024, 0.024$] & [ $-0.029, 0.029$] & [ $-0.015, 0.015$] & [ $-0.017, 0.017$]\\
   & \dkg & [ $-0.099, 0.14$] & [ $-0.13, 0.17$] & [ $-0.058, 0.063$] & [ $-0.067, 0.073$]\\
   & \lambdag & [ $-0.084, 0.084$] & [ $-0.10, 0.10$] & [ $-0.042, 0.041$] & [ $-0.049, 0.049$]\\
\hline
   & \dgz & [ $-0.042, 0.064$] & [ $-0.055, 0.073$] & [ $-0.044, 0.048$] & [ $-0.051, 0.054$]\\
   & \dkz & [ $-0.047, 0.068$] & [ $-0.064, 0.083$] & [ $-0.037, 0.040$] & [ $-0.043, 0.047$]\\
    $\lambdaff=5~\tev$ & \lambdaz & [ $-0.026, 0.026$] & [ $-0.032, 0.032$] & [ $-0.020, 0.019$] & [ $-0.023, 0.022$]\\
   & \dkg & [ $-0.10, 0.15$] & [ $-0.14, 0.18$] & [ $-0.077, 0.084$] & [ $-0.089, 0.097$]\\
   & \lambdag & [ $-0.089, 0.089$] & [ $-0.11, 0.11$] & [ $-0.056, 0.056$] & [ $-0.065, 0.065$]\\
\hline
\hline
\end{tabular}
 \caption{The observed and expected 95\% confidence intervals for the aTGC parameters
without the LEP constraint.  The confidence intervals
are computed separately for the \resolved\ and \boosted\ channels, 
and are calculated both for $\lambdaff=\lffval~\tev$ and $\lambdaff=\infty$ (i.e. no form factor).
The confidence intervals for each parameter are calculated while fixing the other parameters to zero.}
\label{tab:limits_aTGCnoconstraint}
\end{center}
\end{table}

\begin{table}
  \begin{center}
\begin{tabular}{ccccc}
  \hline \hline
 Parameter   &  Observed  & Expected  &  Observed  & Expected  \\
             &  \multicolumn{2}{c}{ \resolved } &  \multicolumn{2}{c}{ \boosted } \\
\hline
  \dgz & [ $-0.027, 0.045$] & [ $-0.036, 0.051$] & [ $-0.021, 0.024$] & [ $-0.024, 0.027$]\\
  \dkg & [ $-0.11, 0.13$] & [ $-0.15, 0.16$] & [ $-0.061, 0.064$] & [ $-0.071, 0.075$]\\
  \lambdaz = \lambdag & [ $-0.022, 0.022$] & [ $-0.027, 0.026$] & [ $-0.013, 0.013$] & [ $-0.015, 0.015$]\\
\hline
\hline
\end{tabular}
  \caption{The observed and expected 95\% confidence intervals for the aTGC parameters
in the LEP-constraint scenario with $\lambdaff=\infty$,  computed separately for the \resolved\ and \boosted\ channels.
The confidence intervals for each parameter are calculated while fixing the other parameters to zero.}
\label{tab:limits_aTGClep}
\end{center}
\end{table}

\begin{figure}[htbp]
\begin{center}
\subfigure[]{
 \includegraphics[width=0.47\textwidth,keepaspectratio]{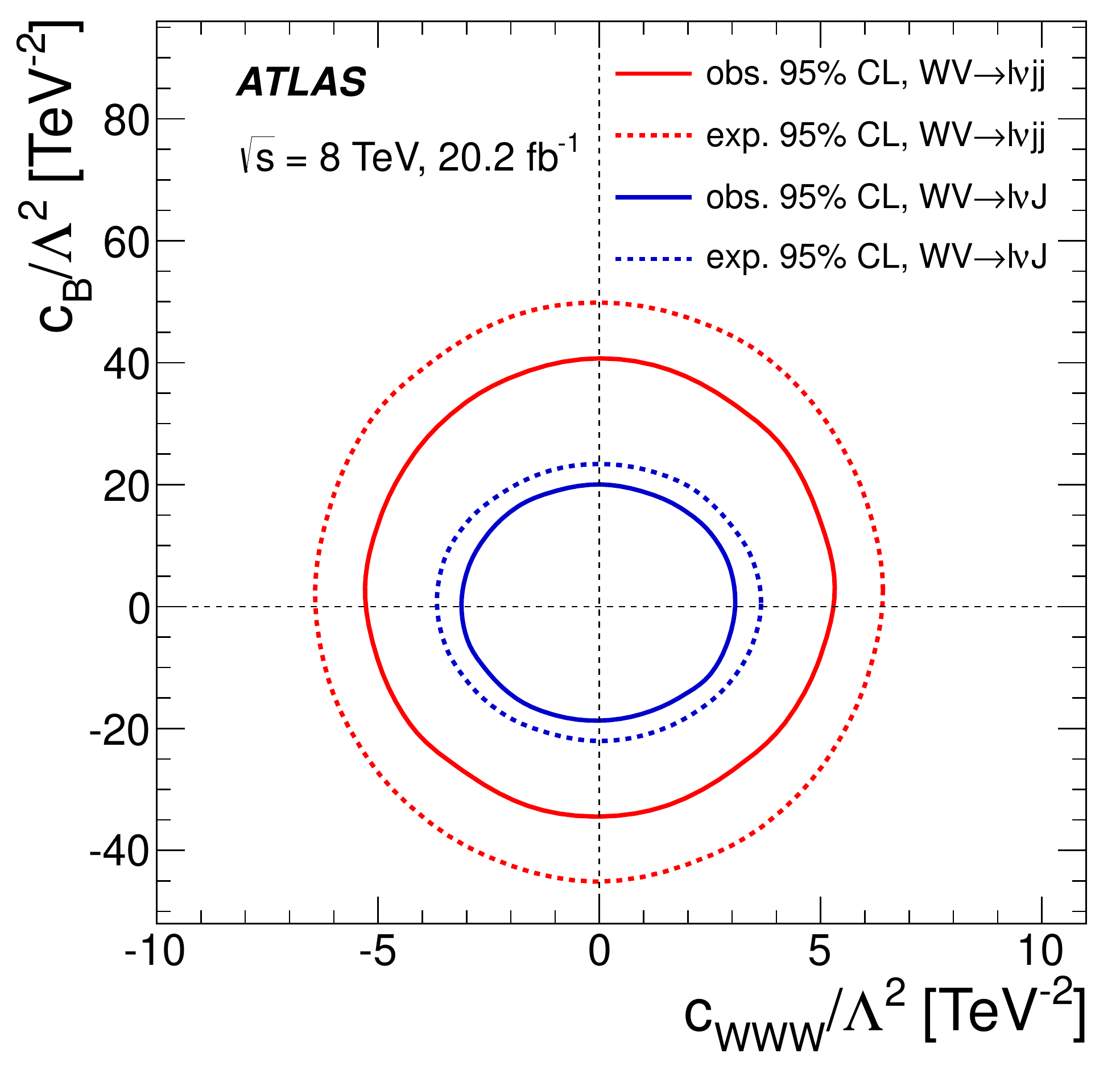}
}
\subfigure[]{
 \includegraphics[width=0.47\textwidth,keepaspectratio]{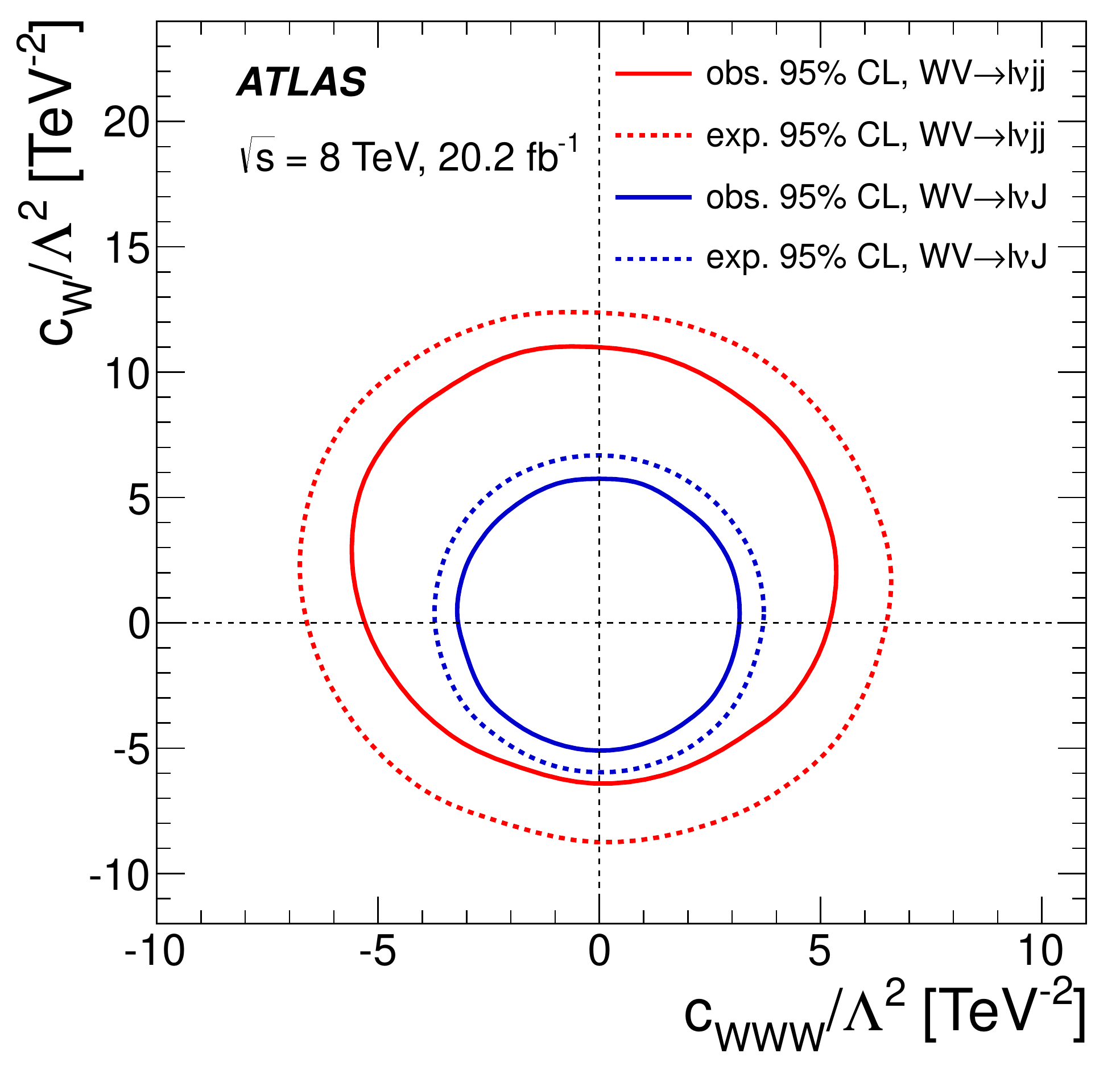}
}
\subfigure[]{
 \includegraphics[width=0.47\textwidth,keepaspectratio]{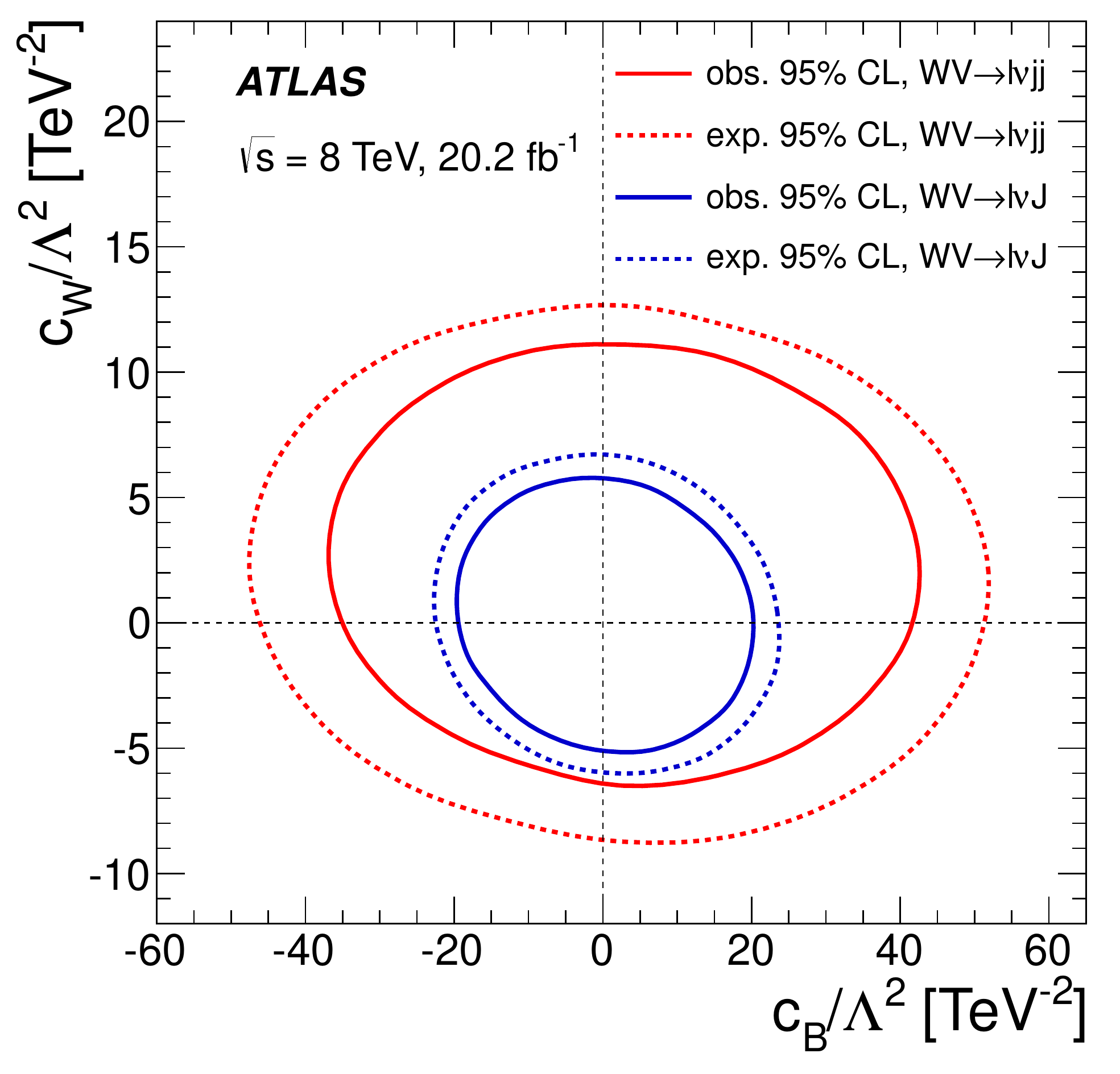}
}
\caption{
The 95\% confidence-level regions for combinations of two EFT parameters.
(a)  \cwww\ and \cb, (b) \cwww\ and \cw, (c) \cb\ and \cw.   The expected
and observed confidence regions are shown for the \resolved\ channel (outer
contours) and the \boosted\ channel (inner contours). 
When computing the confidence regions for two
parameters, the third EFT parameter is held fixed to zero.
}
\label{fig:EFT2dLimits}
\end{center}
\end{figure}

The observed and expected confidence intervals for the EFT parameters
are shown in Table~\ref{tab:limits_EFT}, separately for the \resolved\ and
\boosted\ selections.  Confidence regions for combinations of two
EFT parameters are shown in Figure~\ref{fig:EFT2dLimits}; for each 
combination the third EFT parameter is held fixed to zero.
Although the constraints from the \resolved\ channel are less stringent than those
from the \boosted\ channel, they probe a complementary phase space.
The sensitivity of the  \boosted\ channel 
is similar to
the most sensitive previous analyses to publish constraints on 
these parameters~\cite{STDM-2013-07,STDM-2014-02,CMS-SMP-14-016,Sirunyan:2017bey}.
The \boosted\ channel probes a similar phase space to Ref.~\cite{Sirunyan:2017bey}; 
these analyses benefit from their ability to reconstruct high-$\pT$ $\Vqq$ decays.

\begin{table}
  \begin{center}
\begin{tabular}{ccccc}
  \hline \hline
 Parameter   &  Observed [$\tevminSq$]  & Expected  [$\tevminSq$]  &  Observed [$\tevminSq$] & Expected [$\tevminSq$]  \\
             &  \multicolumn{2}{c}{ \resolved } & \multicolumn{2}{c}{ \boosted } \\
\hline
  \cwww & [ $-5.3, 5.3$] & [ $-6.4, 6.3$] & [ $-3.1, 3.1$] & [ $-3.6, 3.6$]\\
  \cb & [ $-36, 43$] & [ $-45, 51$] & [ $-19, 20$] & [ $-22, 23$]\\
  \cw & [ $-6.4, 11$] & [ $-8.7, 13$] & [ $-5.1, 5.8$] & [ $-6.0, 6.7$]\\
\hline
\hline
\end{tabular}
 \caption{The observed and expected 95\% confidence intervals for the EFT parameters.
The parameters are given in units of $\tevminSq$.  
The  confidence intervals for each parameter are calculated while fixing the other parameters to zero.}
\label{tab:limits_EFT}
\end{center}
\end{table}

%-------------------------------------------------------------------------------

\FloatBarrier 

%-------------------------------------------------------------------------------
\section{Conclusion}
\label{sec:conclusion}
The production of $\WVlnuqq$, with $V$ being a $W$ or $Z$ boson, is measured using $\ourLumiNoError$ of $pp$ collisions at $8~\tev$ at the LHC with the ATLAS detector.
The measurements focus on $WV$ production where the bosons have large transverse
momentum.  Fiducial cross-sections for the $\WVlnuqq$ process are measured in two different,
but partially overlapping, phase spaces.  

The first phase space, denoted \resolved, targets
a hadronically decaying $V$ boson whose decay products can be distinguished as two $R=0.4$ jets.
In this phase space, the $\WVlnuqq$ process is measured with a significance 
of $\resultSignifResolvedObs$, and
the fiducial cross-section is measured to be $\resultSigmafidResolvedObs$, in agreement
with the \MCatNLO\ prediction of $\resultSigmafidResolvedTheory$.

The second phase space, denoted \boosted, contains a single $R=1.0$ jet consistent with
the collimated decay products of a high-$\pT$ $V$ boson.  The $WV$ process is measured 
with a significance of $\resultSignifBoostedObs$ in this phase space.  The fiducial cross-section for this
phase space is measured to be  $\resultSigmafidBoostedObs$, consistent
with the \MCatNLO\ prediction of $\resultSigmafidBoostedTheory$.

The events are also used to search for new physics modifying triple gauge-boson
vertices, which could lead to enhancements of the cross-section at high $\pT$ of the bosons.
No evidence is found for new physics, and $95\%$ confidence intervals are computed for 
anomalous coupling parameters.  The constraints on new physics are also interpreted in terms of
an effective field theory.  The \boosted\ channel is found to be significantly more sensitive to
the new-physics parameters than the \resolved\ channel, which demonstrates the power 
of large-radius jet substructure techniques. The constraints  from this analysis
on the new physics parameters are comparable to the previous most stringent constraints from other diboson analyses.

%-------------------------------------------------------------------------------

%-------------------------------------------------------------------------------
\section*{Acknowledgements}
%-------------------------------------------------------------------------------

% Acknowledgements for papers with collision data
% Version 06-March-2017

% Standard acknowledgements start here
%----------------------------------------------
We thank CERN for the very successful operation of the LHC, as well as the
support staff from our institutions without whom ATLAS could not be
operated efficiently.

We acknowledge the support of ANPCyT, Argentina; YerPhI, Armenia; ARC, Australia; BMWFW and FWF, Austria; ANAS, Azerbaijan; SSTC, Belarus; CNPq and FAPESP, Brazil; NSERC, NRC and CFI, Canada; CERN; CONICYT, Chile; CAS, MOST and NSFC, China; COLCIENCIAS, Colombia; MSMT CR, MPO CR and VSC CR, Czech Republic; DNRF and DNSRC, Denmark; IN2P3-CNRS, CEA-DSM/IRFU, France; SRNSF, Georgia; BMBF, HGF, and MPG, Germany; GSRT, Greece; RGC, Hong Kong SAR, China; ISF, I-CORE and Benoziyo Center, Israel; INFN, Italy; MEXT and JSPS, Japan; CNRST, Morocco; NWO, Netherlands; RCN, Norway; MNiSW and NCN, Poland; FCT, Portugal; MNE/IFA, Romania; MES of Russia and NRC KI, Russian Federation; JINR; MESTD, Serbia; MSSR, Slovakia; ARRS and MIZ\v{S}, Slovenia; DST/NRF, South Africa; MINECO, Spain; SRC and Wallenberg Foundation, Sweden; SERI, SNSF and Cantons of Bern and Geneva, Switzerland; MOST, Taiwan; TAEK, Turkey; STFC, United Kingdom; DOE and NSF, United States of America. In addition, individual groups and members have received support from BCKDF, the Canada Council, CANARIE, CRC, Compute Canada, FQRNT, and the Ontario Innovation Trust, Canada; EPLANET, ERC, ERDF, FP7, Horizon 2020 and Marie Sk{\l}odowska-Curie Actions, European Union; Investissements d'Avenir Labex and Idex, ANR, R{\'e}gion Auvergne and Fondation Partager le Savoir, France; DFG and AvH Foundation, Germany; Herakleitos, Thales and Aristeia programmes co-financed by EU-ESF and the Greek NSRF; BSF, GIF and Minerva, Israel; BRF, Norway; CERCA Programme Generalitat de Catalunya, Generalitat Valenciana, Spain; the Royal Society and Leverhulme Trust, United Kingdom.

The crucial computing support from all WLCG partners is acknowledged gratefully, in particular from CERN, the ATLAS Tier-1 facilities at TRIUMF (Canada), NDGF (Denmark, Norway, Sweden), CC-IN2P3 (France), KIT/GridKA (Germany), INFN-CNAF (Italy), NL-T1 (Netherlands), PIC (Spain), ASGC (Taiwan), RAL (UK) and BNL (USA), the Tier-2 facilities worldwide and large non-WLCG resource providers. Major contributors of computing resources are listed in Ref.~\cite{ATL-GEN-PUB-2016-002}.
%----------------------------------------------

\printbibliography

\newpage
% ATLAS Collaboration author list
% Data extracted on 08-Jun-2017 for paper reference STDM-2015-23
%\documentclass[11pt]{article}
%\usepackage{a4wide}\begin{document}
\begin{flushleft}
{\Large The ATLAS Collaboration}

\bigskip

M.~Aaboud$^\textrm{\scriptsize 137d}$,
G.~Aad$^\textrm{\scriptsize 88}$,
B.~Abbott$^\textrm{\scriptsize 115}$,
O.~Abdinov$^\textrm{\scriptsize 12}$$^{,*}$,
B.~Abeloos$^\textrm{\scriptsize 119}$,
S.H.~Abidi$^\textrm{\scriptsize 161}$,
O.S.~AbouZeid$^\textrm{\scriptsize 139}$,
N.L.~Abraham$^\textrm{\scriptsize 151}$,
H.~Abramowicz$^\textrm{\scriptsize 155}$,
H.~Abreu$^\textrm{\scriptsize 154}$
% ...
\bigskip
\\
% ...
$^{12}$ Institute of Physics, Azerbaijan Academy of Sciences, Baku, Azerbaijan\\
% ...
$^{88}$ CPPM, Aix-Marseille Universit{\'e} and CNRS/IN2P3, Marseille, France\\
% ...
$^{115}$ Homer L. Dodge Department of Physics and Astronomy, University of Oklahoma, Norman OK, United States of America\\
% ...
$^{119}$ LAL, Univ. Paris-Sud, CNRS/IN2P3, Universit{\'e} Paris-Saclay, Orsay, France\\
% ...
$^{137}$ $^{(a)}$ Facult{\'e} des Sciences Ain Chock, R{\'e}seau Universitaire de Physique des Hautes Energies - Universit{\'e} Hassan II, Casablanca; $^{(b)}$ Centre National de l'Energie des Sciences Techniques Nucleaires, Rabat; $^{(c)}$ Facult{\'e} des Sciences Semlalia, Universit{\'e} Cadi Ayyad, LPHEA-Marrakech; $^{(d)}$ Facult{\'e} des Sciences, Universit{\'e} Mohamed Premier and LPTPM, Oujda; $^{(e)}$ Facult{\'e} des sciences, Universit{\'e} Mohammed V, Rabat, Morocco\\
% ...
$^{139}$ Santa Cruz Institute for Particle Physics, University of California Santa Cruz, Santa Cruz CA, United States of America\\
% ...
$^{151}$ Department of Physics and Astronomy, University of Sussex, Brighton, United Kingdom\\
% ...
$^{154}$ Department of Physics, Technion: Israel Institute of Technology, Haifa, Israel\\
$^{155}$ Raymond and Beverly Sackler School of Physics and Astronomy, Tel Aviv University, Tel Aviv, Israel\\
% ...
$^{161}$ Department of Physics, University of Toronto, Toronto ON, Canada\\
% ...
$^{a}$ Also at Department of Physics, King's College London, London, United Kingdom\\
$^{b}$ Also at Institute of Physics, Azerbaijan Academy of Sciences, Baku, Azerbaijan\\
$^{c}$ Also at Novosibirsk State University, Novosibirsk, Russia\\
$^{d}$ Also at TRIUMF, Vancouver BC, Canada\\
$^{e}$ Also at Department of Physics {\&} Astronomy, University of Louisville, Louisville, KY, United States of America\\
% ...
$^{*}$ Deceased
\end{flushleft}

%\end{document}
% Created with xml2latex.py

\clearpage
\appendix
%\part*{Auxiliary material}
%\addcontentsline{toc}{part}{Auxiliary material}
%-------------------------------------------------------------------------------
%\input{Auxiliary}

\end{document}